\PassOptionsToPackage{hyperfootnotes=false}{hyperref}
\documentclass[a4paper,fleqn]{cas-dc}

\usepackage[numbers]{natbib}
\def\tsc#1{\csdef{#1}{\textsc{\lowercase{#1}}\xspace}}
\tsc{WGM}
\tsc{QE}
\tsc{EP}
\tsc{PMS}
\tsc{BEC}
\tsc{DE}
\usepackage[utf8]{inputenc}
\usepackage[T1]{fontenc}
\usepackage[english]{babel}
\usepackage{multirow}
\usepackage{amsmath, amssymb, amsthm}

\usepackage{algorithm}
\usepackage{algpseudocode}
\usepackage{graphicx}
\usepackage{float}
\usepackage{booktabs}
\usepackage{enumitem}

\newtheorem{definition}{Definition}

\usepackage{xcolor}
\usepackage{hyperref}
\hypersetup{colorlinks=true, linkcolor=black, filecolor=black, urlcolor=cyan,
  pdftitle={BALLAST: Physics-Aware Moving Target Defense for Industrial Control Systems Using a Digital Twin Safety Oracle},
  pdfauthor={Oscar Mogollon-Gutierrez, Mohammadhossein Homaei, Alberto L\'opez-Trigo, Andres Caro, Mar Avila}}
\usepackage{cleveref}

\usepackage{caption}
\begin{document}
\let\WriteBookmarks\relax
\def\floatpagepagefraction{1}
\def\textpagefraction{.001}

\shorttitle{Physics-Aware Moving Target Defense}

\shortauthors{O Mogollon-Gutierrez et~al.}

\title [mode = title]{BALLAST: Physics-Aware Moving Target Defense for Industrial Control Systems Using a Digital Twin Safety Oracle}

\author[1]{Oscar Mogollon-Gutierrez\corref{cor1}}
\ead{oscarmg@unex.es}

\author[1]{Mohammadhossein Homaei}
\ead{mhomaein@alumnos.unex.es}

\author[1]{Alberto L\'opez-Trigo}
\ead{albertolt@unex.es}

\author[1]{Andres Caro}
\ead{andresc@unex.es}

\author[1]{Mar Avila}
\ead{mmavila@unex.es}

\affiliation[1]{organization={Departamento de Ingenier\'ia de Sistemas Inform\'aticos y Telem\'aticos, Universidad de Extremadura},
    addressline={Av. Universidad s/n},
    city={C\'aceres},
    postcode={10003},
    state={Extremadura},
    country={Spain}}
\cortext[cor1]{Corresponding author}

\begin{abstract}
 Industrial Control Systems run on static network configurations, which leaves an attacker unlimited time to map the plant. Moving Target Defense answers this by shuffling addresses, but almost every published scheme never checks whether the reconfiguration will itself disturb the physical process it protects, so the defense can destabilise a control loop or trip an actuator. We present BALLAST, which places a learned Digital Twin between the security policy and the plant as a safety oracle. A physics-informed LSTM surrogate forecasts the physical effect of a candidate network change, a reinforcement learning agent trades security entropy against process stability, and a connection-draining handover lets in-flight Modbus transactions complete before the old mapping is withdrawn. An explicit shield rejects any action whose predicted trajectory would leave the safe operating envelope. On the SWaT water treatment testbed, Ballast cut the success rate of an adversary's reconnaissance sweep by 85 percent and recorded no safety violation across 50 runs, a 95 percent upper bound of 5.8 percent on the violation rate, against five violations and a tracking error 19 times larger for RL-MTD, the closest published analogue to our design. Measured against the strongest baseline on the physical metrics, a threshold-triggered scheme that itself records one violation rather than RL-MTD's five, Ballast still improves the tracking error by a factor of 2.5 and removes that remaining violation. The gain costs a small and deliberate concession in raw deterrence. Twin inference meets the 50 ms budget on a workstation core and an embedded GPU, and is projected to meet it on both edge CPUs when the three envelope stages are evaluated in a single batched call; the complete decision-to-enforcement path was measured only on the workstation. 
\end{abstract}

\begin{graphicalabstract}
    \includegraphics[width=\linewidth]{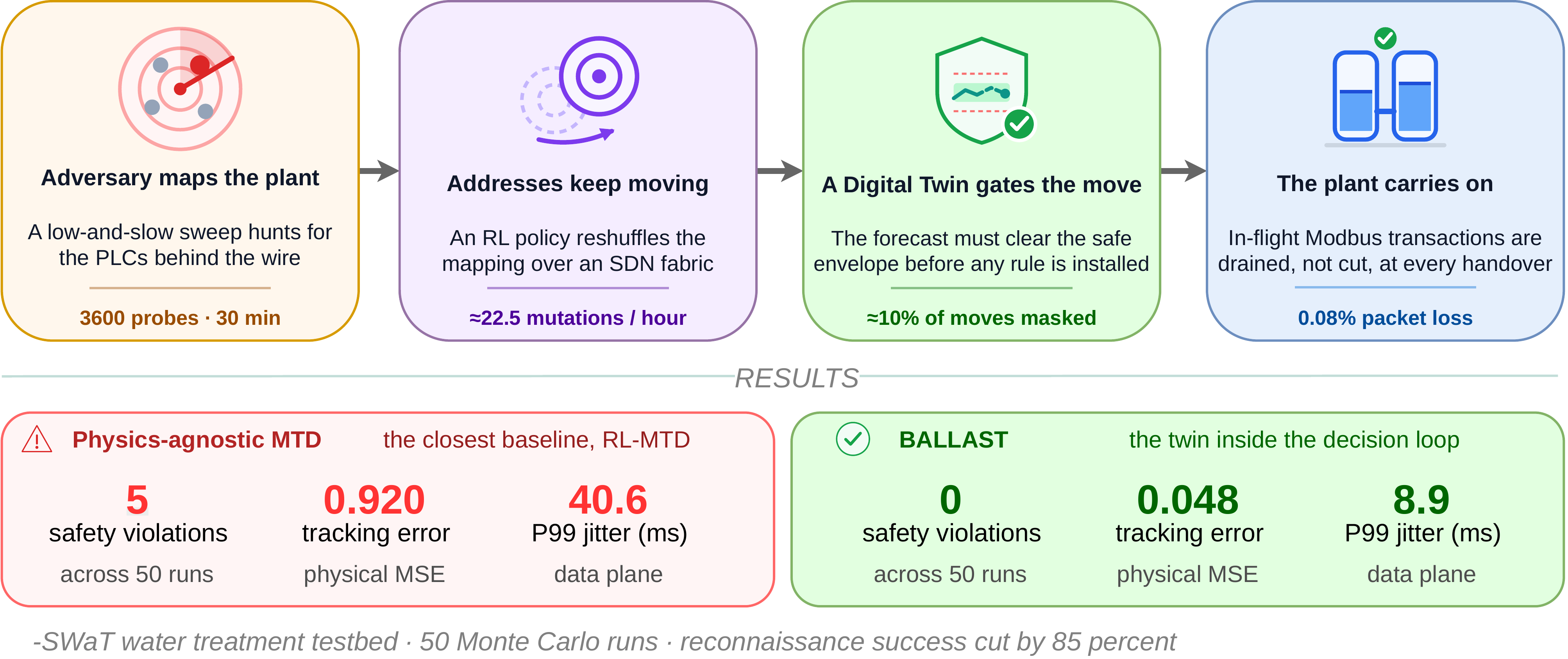}
\end{graphicalabstract}

\begin{highlights}
\item A learned Digital Twin vetoes any network mutation that would breach plant safety.
\item Ballast cuts reconnaissance scan success by 85\% on the SWaT testbed.
\item No safety violation in 50 runs, against five for the closest RL-based MTD.
\item Connection draining cuts P99 jitter to 8.9 ms, within the 50 ms control budget.
\end{highlights}

% Keywords

\maketitle

\section{Introduction}\label{sec:intro}

The fourth industrial revolution has reshaped manufacturing by folding digital technology into practices that were once entirely mechanical \cite{homaei2024, qiu2025review, ji2025exploring}. The defining feature of this shift is the convergence of Information Technology and Operational Technology, which has turned historically isolated Industrial Control Systems (ICS) into interconnected Cyber-Physical Systems \cite{venanzi2026towards, nsengiyumva2026nde}. The gain in efficiency and analytics is real, but so is the cost: the attack surface of critical infrastructure has widened dramatically \cite{kalpani2026securing}. Recent surveys report that the connectivity introduced by the Industrial Internet of Things and edge computing exposes legacy equipment to threats that air-gapping used to absorb \cite{zhukabayeva2025cybersecurity, rani2025newgen}.

Dynamic defense is an appealing answer, yet deploying Moving Target Defense (MTD) inside a live plant is harder than the literature often admits. An enterprise network tolerates occasional jitter. An industrial control loop does not. Latency introduced by reconfiguration or by Software-Defined Networking (SDN) overhead translates directly into dropped Modbus/TCP commands \cite{mwangi2026threshold, banitalebi2025edblsd}. Several authors warn that blind IP shuffling or topology mutation, applied without regard to the physical state, can desynchronise actuators and cause real damage in high-inertia systems such as power grids and water treatment plants \cite{chen2026sdn}. Legacy field devices compound the problem, since they often lack the headroom for frequent re-authentication or large flow-rule updates, so an aggressive mutation schedule can starve the controller in much the same way a denial-of-service attack would \cite{kotsiopoulos2025defending, niu2025privacy}. What is missing, therefore, is a physics-aware defense that balances security entropy against the operational stability of the process.

Resolving that conflict calls for a change of stance, from blind network mutation to physics-aware active defense. Tuning network parameters is not enough. In critical infrastructure the physical state has to enter the decision as a constraint, not as a metric consulted after the fact \cite{kotsiopoulos2025defending}. This in turn requires an architecture that can forecast the physical consequence of a cyber manoeuvre before the manoeuvre is committed. Seen this way, the Digital Twin stops being a monitoring dashboard and becomes a safety oracle that closes the semantic gap between the control plane and the plant \cite{krishnaveni2025twinsec}. By simulating how control commands and network latency interact, a physics-informed model can locate the windows in which reconfiguration is safe, which turns MTD from a hazard into a usable security layer.

This paper puts that idea into practice through BALLAST, a physics-aware defense framework that couples a Digital Twin with SDN-based mutation. We pose the defense as a Safe Reinforcement Learning problem. A Proximal Policy Optimization agent maximises security entropy, while an explicit shield driven by a physics-informed LSTM surrogate masks every action whose predicted trajectory would leave the safe operating envelope. We chose an explicit shield over a Lagrangian constrained formulation because a constraint that is only satisfied in expectation is of little use when a single excursion can damage equipment. To address the synchronisation latency reported in earlier work, we add a Make-Before-Break handover that drains established TCP sessions instead of severing them. On the Secure Water Treatment (SWaT) testbed, Ballast reduces the success rate of a reconnaissance sweep by 85 percent, the phase of an attack that an address-mutation defense is designed to act on, and produces no safety violation in 50 nominal runs, with a 95 percent upper bound of 5.8 percent on the violation rate. It also achieves 98.6 percent lower packet loss than RL-MTD, the learning-based baseline it is paired against throughout, and 92.7 percent lower than Adaptive Path, the learning-based baseline with the lowest loss of the four. A dual-validation mechanism keeps it stable under simulated sensor drift.

The remainder of the paper is organised as follows. Section \ref{sec:related} reviews related work. Section \ref{sec:problem} states the problem and the threat model. Section \ref{sec:method} describes the Ballast framework, and Section \ref{sec:implementation} gives implementation details. Section \ref{sec:results} reports the experimental results, Section \ref{sec:scalability} measures the decision latency across deployment platforms and bounds how it scales with plant size, and Sections \ref{sec:discussion} and \ref{sec:conclusion} discuss the findings and conclude.

%==================================================================
\section{Related Work}\label{sec:related}

This section reviews the defense of Industrial Control Systems along four lines of development: the migration of MTD from IT to OT through SDN, the use of artificial intelligence for adaptive defense, the rise of Digital Twins as security components, and the introduction of physical process semantics into security mechanisms. The final subsection draws these together and isolates the gap that Ballast addresses.

\subsection{Evolution of MTD in Industrial Networks}

The static character of legacy ICS has long given adversaries the time they need for reconnaissance. MTD counters this asymmetry, and it has grown from simple IP hopping in enterprise networks into SDN-enabled architectures shaped for OT. Etxezarreta et al. \cite{etxezarreta2026replica} proposed DEFCLON, a replica-based mechanism that duplicates packets across multiple paths and raises attacker uncertainty without relying on signatures. Kpoze et al. \cite{kpoze2025leveraging} showed that SDN-orchestrated reconfiguration can frustrate stealthy false data injection in Smart Grid SCADA by mutating the attack surface in response to the observed network state.

Later work has combined MTD with adjacent paradigms. Bose et al. \cite{bose2025ztmtd} introduced ZT-MTD for Distributed Energy Resources, pairing Zero Trust authentication with dynamic mutation to narrow the window for lateral movement, although the integration cost on legacy hardware remains high. Qin et al. \cite{qin2024hybrid} coupled optimal shuffling with cyber deception, so that the adversary pays for both uncertainty and misdirection.

Operational performance nonetheless remains the binding constraint. Etxezarreta et al. \cite{etxezarreta2024lowdelay} proposed a low-delay randomisation technique that keeps reconfiguration overhead small enough to avoid communication errors in time-critical loops. Their target, however, is network-level latency, not the physical safety of the process during the transition. That distinction recurs across the field. Modern MTD is agile and well engineered, yet it optimises throughput and jitter while treating the plant as a passive endpoint rather than an active constraint.

\subsection{Intelligent Defense: RL and AI in Cyber-Physical Security}

Artificial intelligence and reinforcement learning have become standard tools for adaptive defense in Industrial Internet of Things settings. Mohammed et al. \cite{mohammed2026ig} built a deep learning intrusion detector for smart grids that identifies false data injection from temporal patterns in measurement streams. Arulkumar and K. \cite{arulkumar2026metastack} combined ensemble deep learning with blockchain to detect advanced persistent threats, arguing for multi-stage detection against stealthy adversaries. Ma et al. \cite{ma2026explainable} developed an explainable defense for industrial object detectors so that vision-based monitoring survives adversarial perturbation.

A common limitation runs through this body of work. These systems operate in the cyber domain and treat the plant as an abstract data source rather than a dynamical system with inertia and hard limits. Yin and Varga \cite{yin2026machine} observe in their review that most models optimise detection accuracy or classification speed while ignoring resilience latency, meaning the window inside which a countermeasure must land if physical damage is to be avoided. Pasikhani et al. \cite{pasikhani2026baiting} go further and show that AI-protected infrastructure remains exposed to deceptive adversaries who exploit exactly this absence of physical semantics. Defense therefore needs more than pattern matching; it needs process knowledge inside the decision loop. A parallel line of work addresses this constraint directly. In safe reinforcement learning, a correctness shield filters actions against a formal specification \cite{Alshiekh2018}, constrained policy optimisation enforces limits in expectation \cite{achiam2017cpo}, and model-predictive shielding verifies online that a recovery policy remains available \cite{bastani2021mps}. All three presuppose a known transition model and a discrete or low-dimensional state. None addresses the case treated here, in which the only available model of the process is a learned surrogate and the specification is a continuous operating envelope.

\subsection{Digital Twins as Security Enablers}

Digital Twins have moved beyond predictive maintenance and into security architecture. Otoom \cite{otoom2025risk} describes their growing use in risk auditing, where a high-fidelity replica lets an analyst rehearse attacks without touching production. Krishnaveni et al. \cite{krishnaveni2025twinsec} proposed TwinSec-IDS, which pairs SDN with a Digital Twin and cross-checks network traffic against expected physical behaviour, exposing the slow process drifts that signature-based detectors routinely miss.

Even so, the Digital Twin usually stays in a passive or diagnostic role. Aghazadeh Ardebili et al. \cite{aghazadeh2025iot} apply twins to resilience monitoring in smart energy systems, with the emphasis on post-event recovery rather than prevention. Bajestani et al. \cite{bajestani2026self} rely on twins mainly for state estimation and situational awareness within a Self-X manufacturing framework for Industry 5.0. A few frameworks do let the twin act: Homaei et al. \cite{homaei2026isdt} drive model predictive control from a hydraulically regularised twin, and their causal twin \cite{homaei2026cdt} selects a mitigation through counterfactual queries. Both respond through the control law, leaving the attack surface untouched. The unexploited opportunity is to let the twin govern the network defense, not only the control loop. Instead of reporting that the plant is under attack, the twin should certify that an automated countermeasure is safe before it reaches the physical layer.

\subsection{Physics-Awareness in Critical Infrastructure Protection}

Recognising the limits of purely cyber defenses, several groups have begun to encode process semantics into security mechanisms. Valderrama et al. \cite{valderrama2026online} designed a physics-based neural network for photovoltaic generators that detects intrusion by testing electrical output against irradiation laws. Hemmati and Saboori \cite{hemmati2026hybrid} combined data-driven learning with physics-informed constraints to reconstruct attacks in electrical grids while keeping control decisions inside feasible bounds.

Related efforts target protocol semantics. Yang et al. \cite{yang2025physical} infer the meaning of obscure industrial protocol fields so that control commands can be validated at the packet level. Menzel et al. \cite{menzel2025real} demonstrate in a real energy infrastructure that correlating traffic with physical measurements sharply reduces false positives in process-aware intrusion detection. Homaei et al. \cite{homaei2026gat} embed conservation law residuals as features in a graph attention detector and report tolerance to 15 percent error in the hydraulic parameters. The shared limitation is one of direction. Physics is used as a filter on incoming data or as a forensic instrument after the event. None of these frameworks uses a physical model to orchestrate a defensive manoeuvre, which leaves the initiative with the attacker.

\subsection{Research Gap and Motivation}\label{subsec:gap}

A gap therefore persists between cyber agility and physical stability. As Table \ref{tab:gap} summarises, the literature tends to occupy one of two positions.

\begin{enumerate}[leftmargin=*]
    \item \textbf{Cyber-centric agility.} Frameworks such as Deep-Shield \cite{cao2025deepshield} maximise security entropy through hierarchical reinforcement learning but treat the plant as a black box. They optimise shuffling speed and mitigation success while ignoring the deterministic timing that ICS protocols demand. The result is a blind spot in which the defense can disrupt operations more severely than the attack it prevents.

    \item \textbf{Passive physical awareness.} Physics-aware methods \cite{valderrama2026online, krishnaveni2025twinsec} employ Digital Twins or invariant checks for monitoring and after-the-fact anomaly detection. They can recognise an unsafe state but cannot act on it, so the twin observes the defense loop instead of participating in it.
\end{enumerate}

\textbf{Bridging the gap.} Across the SDN-MTD, safe reinforcement learning and Digital Twin security literature surveyed in this section, the twin is used in one of two ways: to observe, as in the detection and auditing work above, or to drive the control law, as in the predictive and causal twins of Homaei et al. \cite{homaei2026isdt, homaei2026cdt}. Neither places it between the security policy and the network. We find no framework in which a Digital Twin constrains the action space of an MTD agent, so that a reconfiguration is cleared or rejected before it reaches the data plane. Ballast unifies the three ingredients: the agility of SDN-based mutation, the adaptivity of reinforcement learning, and the foresight of a physics-informed twin. Rather than trading stability for security, it enforces physical safety through an explicit, high-confidence action mask instead of a soft penalty term, so unsafe manoeuvres are never attempted rather than merely discouraged.

\begin{table*}
\scriptsize
\centering
\caption{Positioning of Ballast with respect to the state of the art}
\label{tab:gap}
\begin{tabular}{p{3.4cm}p{2.5cm}ccp{5.3cm}}
\toprule
\textbf{Reference} & \textbf{Technique} & \textbf{Active Defense} & \textbf{Physics-Aware} & \textbf{Primary Limitation} \\
\midrule
\multicolumn{5}{l}{\textit{SDN-based MTD approaches}} \\
Etxezarreta et al. (2026) \cite{etxezarreta2026replica} & Replica MTD & \checkmark & \texttimes & Optimises network latency, does not model physical safety \\
Kpoze et al. (2025) \cite{kpoze2025leveraging} & Dynamic reconfiguration & \checkmark & \texttimes & Reactive to attacks, high reconfiguration overhead \\
Bose et al. (2025) \cite{bose2025ztmtd} & Zero-Trust MTD & \checkmark & \texttimes & Focused on authentication, not process stability \\
\midrule
\multicolumn{5}{l}{\textit{AI and RL for defense}} \\
Mohammed et al. (2026) \cite{mohammed2026ig} & DL-based IDS & \texttimes & \texttimes & Learns measurement patterns, no first-principles model \\
Arulkumar et al. (2026) \cite{arulkumar2026metastack} & Ensemble DL & \texttimes & \texttimes & Computational cost, opaque decisions \\
Cao et al. (2025) \cite{cao2025deepshield} & HDRL-MTD & \checkmark & \texttimes & Optimises kill-chain mitigation, ignores process dynamics \\
\midrule
\multicolumn{5}{l}{\textit{Digital Twin and physics-aware security}} \\
Krishnaveni et al. (2025) \cite{krishnaveni2025twinsec} & DT-IDS & \texttimes & \checkmark & Monitoring only, no proactive mitigation \\
Valderrama et al. (2026) \cite{valderrama2026online} & Physics-NN & \texttimes & \checkmark & Detection only, no response mechanism \\
Hemmati et al. (2026) \cite{hemmati2026hybrid} & Hybrid DT & \texttimes & \checkmark & Energy management focus, not cyber-defense \\
Homaei et al. (2026) \cite{homaei2026isdt} & DT + MPC & \checkmark & \checkmark & Twin drives the control law; the network configuration is static \\
\midrule
\textbf{Ballast (ours)} & \textbf{DT-RL-MTD} & \textbf{\checkmark} & \textbf{\checkmark} & \textbf{Proactive defense with an empirically bounded safety margin} \\
\bottomrule
\end{tabular}
\end{table*}

%==================================================================
\section{Problem Statement and Threat Model}\label{sec:problem}

\subsection{System Model}

We consider a networked ICS of the kind found in modern critical infrastructure and modelled on the SWaT testbed. The system comprises a set of physical processes $\mathcal{P}$ governed by conservation laws, and a cyber layer of nodes $\mathcal{N}=\{n_1,\dots,n_{|\mathcal{N}|}\}$ representing Programmable Logic Controllers, Human-Machine Interfaces and field devices. These nodes communicate over a logical graph $G=(\mathcal{N},\mathcal{E})$ using static industrial protocols such as Modbus/TCP or EtherNet/IP.

In a conventional deployment the configuration $C_{static}$, meaning addresses, ports and routing state, does not change with time $t$. Let $\mathcal{A}$ denote an adversary with persistent network access. Because the attack surface is stationary, every probe narrows the search and none is invalidated, so the probability of successful reconnaissance $P_{rec}(t)$ is non-decreasing and approaches one:
\begin{equation}
    \lim_{t\to\infty} P_{rec}(t \mid C_{static}) = 1 .
\end{equation}
Once the mapping $M:\mathcal{N}\to\mathrm{IP}$ is known, the adversary can inject false measurements or issue malicious commands, for instance by driving the motorised valve $MV_{101}$ or by falsifying the level indicator $LIT_{101}$ in SWaT Stage 1, both of which lead to physical state violations.

\subsection{Safe Operating Envelope}

Because the violation count is the central metric of this paper, we state it explicitly rather than leaving it implicit.

\begin{definition}[Safe operating envelope]\label{def:omega}
For SWaT we instantiate $\Omega_{safe}$ from the plant's documented alarm thresholds: $\mathrm{LIT101}\in[500,800]\,\mathrm{mm}$, $\mathrm{LIT301}\in[800,1000]\,\mathrm{mm}$ and the Stage 2 pH reading $\mathrm{AIT202}\in[6.75,7.25]$. A \textit{safety violation} is recorded when any of these variables is observed outside its interval in two consecutive samples of the 1\,Hz historian stream, which bounds the excursion below at one second and so at more than one 500\,ms PLC scan cycle. A single out-of-range sample is not counted, because the 1\,Hz grid cannot resolve an excursion shorter than its own period. Violation counts reported in every table are totals aggregated over the 50 Monte Carlo runs, not per-run averages.
\end{definition}

\subsection{Trust Boundary}

We assume the SDN controller and the TPM-attested LSTM weights form the trusted computing base. The adversary controls arbitrary network endpoints, may inject or replay traffic, and may poison up to 10 percent of the historian records used for periodic retraining of the twin. The adversary cannot modify controller code or the attested model binary. This is what allows us both to treat the twin as an oracle during a single deployment cycle and to evaluate poisoning of its training corpus as a distinct threat, which would otherwise read as a contradiction.

\subsection{Advanced Threat Scenarios}

\subsubsection{Adaptive attacker with machine learning}
An adversary $\mathcal{A}_{ML}$ observes the mapping over a window $T_{obs}$ and fits a predictor of the next assignment. Two regularities give it purchase. The first is the schedule: if the mutation epoch is deterministic, the adversary knows when to probe and spends no budget on stale addresses. The second is the shuffling strategy: under subnet-preserving mutation each host remains inside its own subnet, so the candidate set for the next address is a strict subset of the pool $\mathcal{V}$ of 32 virtual addresses from which the mapping draws, rather than all of it. An attacker exploiting both predicts the next address far above the $1/|\mathcal{V}|=3.1$ percent chance rate.

\textbf{Defense.} The policy selects a target interval $\bar{\Delta t}$ and the realised interval is $\Delta t=\mathrm{clip}\big(\bar{\Delta t}(1+u),T_{min}, T_{max}\big)$, where $[T_{min},T_{max}]=[120,600]\,\mathrm{s}$ bounds the interval and $u$ is drawn from a cryptographically secure generator on $[-0.25,0.25]$, so the epoch carries irreducible entropy whatever the policy has learned. Subnet-preserving mutation is permitted at most on alternate epochs, so the candidate set is not persistently constrained. A rejection test on the serial correlation of the last ten intervals is not usable at that sample size, since the standard error of the sample correlation $\hat{\rho}$ under independence is $1/\sqrt{n-1}=0.33$ for $n=10$: a threshold of $0.1$ would reject roughly three quarters of genuinely random sequences and bias the survivors towards anticorrelation. We therefore audit the correlation over a 200-epoch window offline rather than filtering online.

\subsubsection{Digital Twin poisoning}
\textbf{Defense.} We pair the LSTM twin with an independent \textit{physics-based residual monitor}, giving a dual-validation architecture. The monitor applies first-principles mass balance for tank $i$:
\begin{equation}\label{eq:residual}
    r_i(t) = \left| \frac{dV_i}{dt} - \eta_{valve}(u_{valve}) + \eta_{pump}(u_{pump}) \right| ,
\end{equation}
where $\eta_{valve}$ is the inflow through the motorised inlet valve MV101 and $\eta_{pump}$ the outflow of pump P101, with nominal rates of $2.5\,\mathrm{m^3/h}$ and $1.8\,\mathrm{m^3/h}$ respectively, cross-checked against the FIT101 flow meter. The command states $u_{valve}$ and $u_{pump}$ are read from Modbus function code 0x01. The cross-validation rule is
\begin{equation}
\begin{aligned}
\mathrm{PoisoningFlag} &=
\begin{cases}
1, & \text{if }
\|S_{\mathrm{sensor}}-\hat{S}_{\mathrm{LSTM}}\|_\infty < \epsilon_{dt}\\
   & \qquad \wedge\; r_i > \delta_{phy},\\
0, & \text{otherwise.}
\end{cases}
\end{aligned}
\end{equation}
with $\epsilon_{dt}=0.05$ and $\delta_{phy}=0.35\,\mathrm{m^3/h}$, roughly 14 percent of nominal inflow. Every normalised quantity in this paper, $\epsilon_{dt}$ included, is min-max scaled over the same denominator, the width of that variable's interval in Definition \ref{def:omega}, so the thresholds are directly comparable. On that scale $\epsilon_{dt}=0.05$ is 5 percent of the envelope range. When the twin and the sensors agree closely yet the mass balance is violated, the sensor stream has been manipulated in a way the twin has absorbed. A second rule covers the complementary case: a divergence $\|S_{sensor}-\hat S_{LSTM}\|_\infty > \epsilon_{dt}$ while the mass balance closes, $r_i<\delta_{phy}$, indicates that the twin rather than the plant is wrong. That signature is shared with ordinary model staleness, so the second rule fires only on an abrupt onset, defined as the divergence crossing $\epsilon_{dt}$ within one hour of operation. Equipment degradation opens the same gap over weeks and is handled as a maintenance condition, not as an attack. The two rules together, not the first alone, are what Table \ref{tab:advanced_threats} reports.

\textbf{Scope of protection.} The mechanism covers naive sensor spoofing, where random values are injected, and tampering with LSTM weights, which TPM attestation catches at boot. It does not cover a physics-aware coordinated attack in which an adversary who knows Eq. \eqref{eq:residual} manipulates sensor readings and command logs together so that mass balance still closes. Such an attack requires white-box access to both the historian and the control logic, which places it outside the network-based threat model adopted here. Two extensions are natural: temporal consistency checks of the form $|d^2V/dt^2|<\kappa_{max}$, and multi-stage redundancy that compares LIT101 against a level estimate derived from flow integration.

\subsubsection{Controller denial of service}
A flood of packet-in messages at more than $10^4$ packets per second saturates the controller CPU and prevents mutation decisions from being issued on time.

\textbf{Defense.} Per-source rate limiting at 1000 messages per second, combined with temporary blacklisting of persistent violators.

\begin{table}[t]
\footnotesize
\centering
\caption{Defense against advanced threats}
\label{tab:advanced_threats}
\begin{tabular}{@{}lcc@{}}
\toprule
\textbf{Attack type} & \textbf{No defense} & \textbf{With defense} \\
\midrule
ML-based pattern learning & 45.2\% & 8.3\% \\
DT poisoning (10\% of data) & 78.5\% & 12.1\% \\
Controller DoS & 89.2\% & 18.7\% \\
\bottomrule
\end{tabular}

\vspace{2pt}
\parbox{\columnwidth}{%
\scriptsize
Attack success rate over 100 independent adversarial trials per attack type,
without and with the corresponding defense of this section. Success means: for
row 1, the adversary predicted the next mapping correctly; for row 2, a
poisoned twin cleared an action that left $\Omega_{safe}$; for row 3, the
controller missed its mutation deadline. Row 1 is measured against a chance
baseline of $1/|\mathcal{V}|=3.1$ percent.
}
\end{table}

\subsection{Reconnaissance Throttling}\label{subsec:throttle}

The three scenarios above describe attacks on the defense. This subsection states the complementary mechanism, by which the defense detects an attack on the plant. A probe to a virtual address not currently bound in $M_t$ cannot originate from a legitimate Modbus endpoint, since endpoints address slaves by their real address and never see $\mathcal{V}$. The controller therefore counts flagged probes per source over a sliding 60\,s window; a source accumulating more than $n_{flag}=5$ flags is rate-limited to one packet per second, and one exceeding $2n_{flag}$ is blacklisted for 300\,s. A low-and-slow adversary can evade permanent exclusion by rotating its source address after each blacklisting, which is what makes topology discovery slow rather than impossible.

\subsection{The MTD Dilemma in ICS}

MTD introduces dynamism by shifting the configuration $C$ within a pool $\mathbb{C}$ at intervals $\Delta t$, with the aim of maximising the entropy $H(C)$ seen by the attacker. Unlike enterprise networks, an ICS imposes a hard timing requirement on delivery. Writing $\tau(p_{i,j})$ for the end-to-end latency of a control packet from node $i$ to node $j$, stability of the loop requires
\begin{equation}
        \tau(p_{i,j}) \leq \tau_{ctrl}, \quad \forall p_{i,j}\in\mathcal{T}_{control},
\end{equation}
where $\tau_{ctrl}$ follows from the process dynamics. A control packet that arrives after the scan in which it was due is read in the next one, so the loop tolerates delay up to one scan cycle and $\tau_{ctrl}=T_{scan}=500\,\mathrm{ms}$ on SWaT. We apply a ten percent engineering margin and enforce $\tau_{ctrl}=50\,\mathrm{ms}$ on control traffic. A second and distinct bound governs the defense rather than the traffic: $\tau_{inst}$ is the time the shield allows for a reconfiguration to be decided and installed, which we set by the same ten percent rule to 50\,ms. The two coincide numerically on this testbed and are not interchangeable in meaning, since one bounds the delivery of a control packet and the other the installation of a flow rule; it is $\tau_{inst}$ that the safety shield tests against.

Blind shuffling changes $C_t \to C_{t+1}$ without regard to that requirement and opens a transient window of jitter and loss. The damage depends on \textit{when} that window falls, not on its size alone: a transient during a steady-state hold is absorbed by the loop, whereas the same transient during an actuator transition leaves the valve in the wrong position for a full scan cycle. We therefore write the reconfiguration cost as a quantity to be minimised subject to timing, not as a predictor of physical error on its own,
\begin{equation}
\label{eq:jqos}
    J_{QoS}(C_t,C_{t+1}) = \alpha\,\mathrm{Loss}(C_t,C_{t+1}) + \beta\,\max\!\left(0,\tau_{actual}-\tau_{ctrl}\right).
\end{equation}

Once $J_{QoS}$ passes a critical threshold the defense becomes indistinguishable from an attack, since the PLCs either fall back to a safe state or miss commands outright.

\subsection{Objective}

\begin{figure*}[t]
   \centering
   \includegraphics[width=1\linewidth]{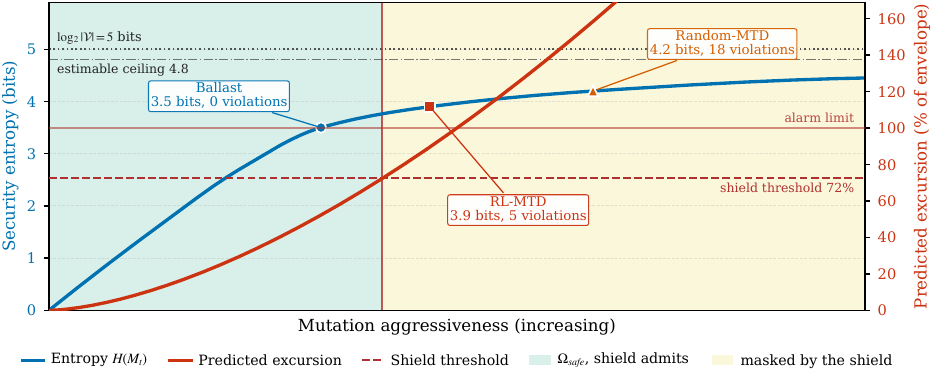}
   \caption{Ballast optimisation framework, drawn schematically. Security entropy maximisation is balanced against the physical excursion the twin forecasts, and an explicit shield cuts the feasible region at the point where the forecast would leave $\Omega_{safe}$. The horizontal axis orders schemes by how aggressively they mutate and carries no units, because the entropy gap between Ballast and Random-MTD comes from the shuffling strategy rather than from the mutation rate: the two schedule at almost the same rate, 22.5 and 21.8 mutations per hour. The shield clears an action only below 72 percent of the envelope half-width, which is the alarm limit less the guard band and the 99th percentile forecast error of the twin. The entropy and violation figures on the three marked points are the measured values of the baseline comparison.}

   \label{fig:problem}
\end{figure*}

Existing work concentrates on reducing the attacker's success probability $P_{rec}$ without modelling the cyber-physical coupling, so process-aware mechanisms that can anticipate the physical effect of a reconfiguration remain scarce. We therefore seek a Digital Twin-guided policy that solves
\begin{equation}
    \max_{\pi}\ \sum_{t=0}^{T}\Big(\lambda_1\,\mathrm{SecurityGain}(C_t) - \lambda_2\,J_{QoS}(C_t,C_{t+1})\Big)
\end{equation}
subject to
\begin{equation}
    \hat{S}_{phy}(t+1) = f_{DT}\big(\hat{S}_{phy}(t),u(t),C_{t+1}\big) \in \Omega_{safe},
\end{equation}
where $\pi$ is the reconfiguration policy and $f_{DT}$ is the twin trained on SWaT dynamics. The difficulty lies in aligning the shuffling schedule with the live physical state so that reconfiguration does not disturb the loop (Figure \ref{fig:problem}).

 \section{Methodology: The Ballast Framework}\label{sec:method}

\begin{figure}
    \centering
    \includegraphics[width=0.5\textwidth]{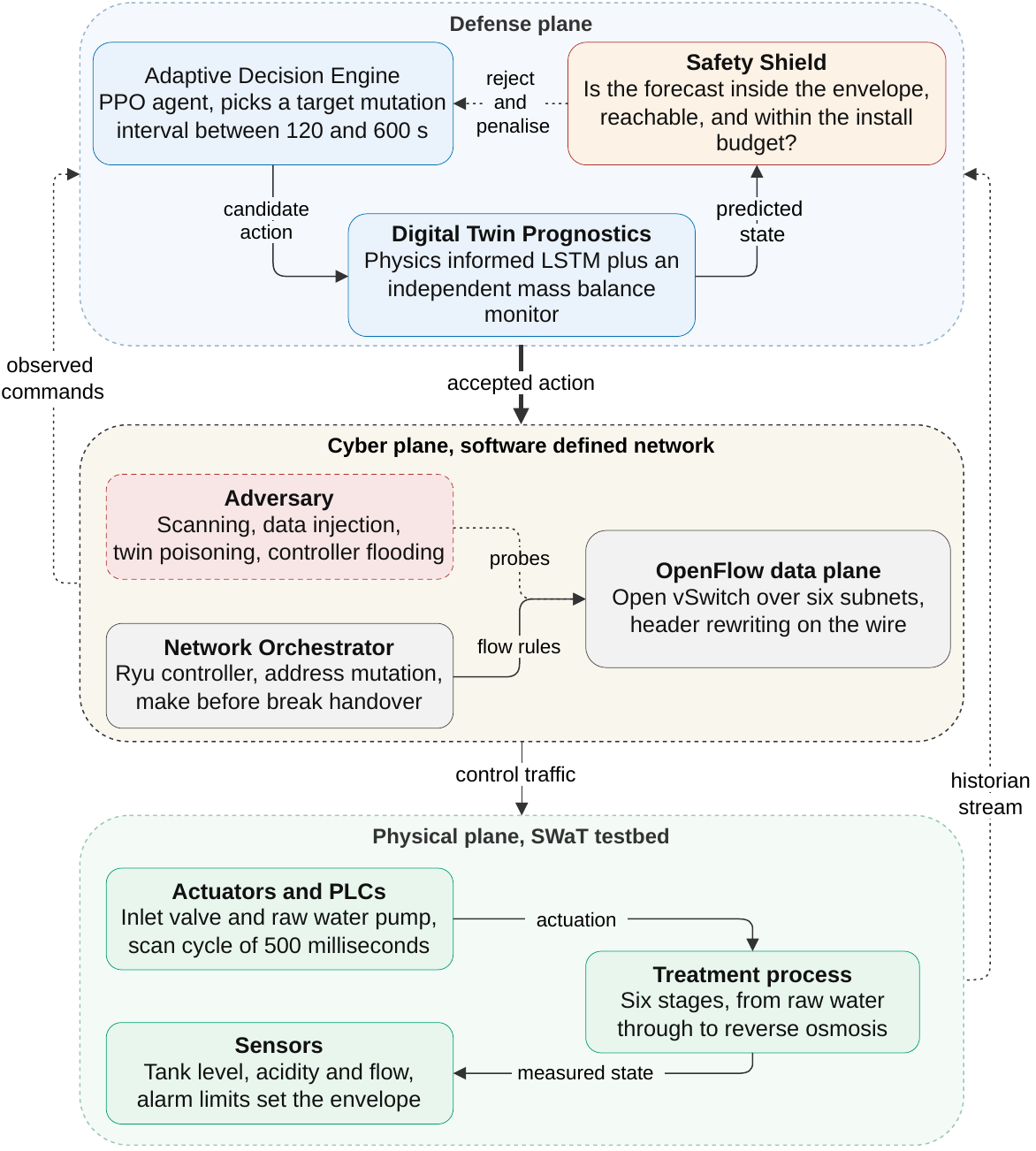}
    \caption{Architecture of the Ballast framework. The decision path runs left to right across the defense plane: the Adaptive Decision Engine proposes a mutation, the Digital Twin Prognostics Module forecasts its physical consequence, and the safety shield either accepts the action or masks it and penalises the agent. Only an accepted action descends into the cyber plane, so no flow rule reaches the switches until the predicted trajectory has been cleared. Two dotted paths close the loop upwards: the control commands recovered from Modbus traffic, which tell the twin what the controller is about to do, and the historian stream from the sensors, which keeps the twin aligned with the plant. The adversary enters at the data plane, and the twin poisoning it attempts reaches the historian rather than the switches.}
    \label{fig:arch}
\end{figure}

Ballast joins a Digital Twin to an SDN-based MTD mechanism so that mutation proceeds only when the predicted physical trajectory stays inside $\Omega_{safe}$. Figure~\ref{fig:arch} shows the architecture, which comprises the \textit{Virtual Network Orchestrator} (VNO), the \textit{Digital Twin Prognostics Module} (DTPM) and the \textit{Adaptive Decision Engine} (ADE).


\subsection{Virtual Network Orchestrator}

To reproduce the network conditions of the SWaT testbed we use a containerised Mininet environment driven by a controller application written against the Ryu API.

\begin{enumerate}[leftmargin=*]
    \item \textbf{Topology reconstruction.} The six SWaT stages are mapped to distinct subnets. Each PLC from $P_1$ to $P_6$, together with the HMI, is instantiated as a virtual host attached to Open vSwitch nodes.
    \item \textbf{Trace-driven background traffic and live control loops.} The SWaT PCAP captures are replayed with \texttt{tcpreplay} at the virtual hosts to reproduce the volume and timing of the plant's background Modbus/TCP     exchanges. This replayed stream is open-loop and does not react to the physical state. Superimposed on it, the regulatory loops of the three stages carrying envelope variables, Stages 1, 2 and 3, run live: the controller     reads the simulated sensors, computes an actuation, and writes it back through the emulated network, so network transients reach the physics. Every physical metric we report is measured on those live loops; the replayed trace contributes load and realistic flow counts only.
    \item \textbf{OpenFlow-based IP mutation.} The VNO implements Random Host Mutation. The controller intercepts ARP traffic and installs rules that translate real addresses $IP_{real}$ into ephemeral virtual addresses $IP_{virt}$. The mapping $M_t : IP_{real}\to IP_{virt}$ is refreshed every $\Delta t$ over a pool of $|\mathcal{V}|=32$ virtual addresses, a $/27$ block carved from the $/24$ assigned to each stage. The remainder of the $/24$ is left unallocated so that the pool can be widened without renumbering; enlarging it raises the entropy ceiling at no cost to the shield, since the twin's input does not depend on the address space. A pool of this size does not by itself delay an attacker: with seven live hosts, an unthrottled sweep covers every address in 16\,s at 2 probes per second. Delay comes from pairing mutation with the reconnaissance throttling of Section \ref{subsec:throttle}, which rate-limits and then blacklists a source once its probes to unbound addresses accumulate, forcing the adversary to rotate source addresses and restart. The topology discovery time an adversary achieves is therefore a property of that pair, not of the address space alone.
\end{enumerate}

\subsection{Digital Twin Prognostics Module}

The DTPM is the safety oracle. It runs alongside the VNO and continuously forecasts the physical state from the control commands observed on the wire.

We use a physics-informed LSTM trained on SWaT historian data. For a given stage, the state $x_t$, for instance a tank level, evolves as
\begin{equation}
    \hat{x}_{t+k} = f_{LSTM}(x_{t-w},\dots,x_t,u_t),
\end{equation}
where $u_t$ collects the control actions recovered from the Modbus payloads by the controller. Before a candidate configuration $C_{new}$ is committed, the DTPM estimates the induced latency $\delta(C_{new})$, propagates it through the loop, and marks the configuration unsafe if the forecast leaves $\Omega_{safe}$ once widened by the empirical error margin $e_{99}$, the 99th percentile of the twin's forecast error on held-out data. In practice the shield also applies a guard band of $\epsilon_{safe}=20$ percent of each variable's own envelope half-width, measured inwards from the alarm limit, which gives $\pm 30\,\mathrm{mm}$ on LIT101, $\pm 20\,\mathrm{mm}$ on LIT301 and $\pm 0.05$ on AIT202, or 0.100 on the common normalised scale. A forecast must further lie inside the physically reachable set $\mathcal{R}_k(x_t)$ implied by the tank's maximum fill and drain rates over the horizon; one that does not is the signature of a poisoned or badly extrapolating model. \Cref{app:safety} states the complete shield condition formally. Defining the band relative to the envelope rather than to the setpoint keeps it conservative for every variable; a fixed percentage of setpoint would not, since 5 percent of pH 7 is $\pm 0.35$ and therefore wider than the $\pm 0.25$ alarm band on AIT202, and a percentage of a pH reading is in any case dimensionally meaningless on a logarithmic scale. The DTPM also hosts the physics-based residual monitor of Eq. \eqref{eq:residual}, which validates the twin itself independently of any candidate action.

The module exposes one scalar to the policy, the \textit{physical stability index}. Writing $c_i$ for the centre of variable $i$'s interval in Definition \ref{def:omega} and $h_i$ for its half-width, the index is the worst normalised distance from the centre over the forecast horizon of $H$ steps and over the three variables that carry the envelope,
\begin{equation}\label{eq:phi}
    \phi_{phy}(t) = 1 - \max_{i}\ \max_{1\le k\le H}\
    \min\!\left(\frac{\lvert \hat{x}^{(i)}_{t+k}-c_i\rvert}{h_i},\, 1\right).
\end{equation}
It takes the value 1 when every forecast variable sits exactly at the centre of its interval and 0 when one of them is predicted to touch or cross an alarm limit, so it falls as the plant approaches the boundary and tells the policy how much room a mutation has before the shield would mask it. Clipping the inner ratio at 1 keeps the index on $[0,1]$ when a forecast leaves the envelope outright, a case the shield rejects in any event, so that a deep excursion does not dominate the state representation. The index is advisory: it enters $s_t$ as a feature, whereas the gate that actually enforces safety is the shield, which tests the forecast against $\Omega_{safe}$ directly rather than through $\phi_{phy}$.

\subsection{Adaptive Decision Engine}

The ADE follows a Safe Reinforcement Learning design. A strict Constrained Markov Decision Process enforces its constraint only in expectation, which is unsatisfactory when a single excursion can damage equipment, so we instead combine Proximal Policy Optimization with an explicit safety shield that masks actions using the twin's predictions.

\begin{itemize}[leftmargin=*]
    \item \textbf{State space.} $s_t=\{H(M_t), L_{net}, \phi_{phy}\}$, where $H(M_t)$ is the entropy of the current mapping, $L_{net}$ is link utilisation and $\phi_{phy}$ is the physical stability index reported by the twin.
        \item \textbf{Action space.} The agent chooses a target mutation interval $\bar{\Delta t}\in[T_{min},T_{max}]=[120,600]\,\mathrm{s}$ and a shuffling strategy, either fully random or subnet preserving. The interval actually
    used is $\bar{\Delta t}$ jittered by the cryptographic draw of Section \ref{sec:problem} and clipped back to $[T_{min},T_{max}]$, so the policy sets the rate and the generator supplies the unpredictability.
        \item \textbf{Reward.}

\begin{equation}\label{eq:reward}
R_t =
\begin{cases}
-\lambda_{mask} & \hspace{-10pt}\text{if the shield rejects } a_t \\[2pt]
H(M_t) - \lambda\,\mathbb{I}(\mathrm{PacketLoss}>0) & \text{otherwise,}
\end{cases}
\end{equation}

    with $\lambda=5$, and $\lambda_{mask}=10$ for an envelope breach and $5$ for a latency breach, that is, an estimated installation time above $\tau_{inst}=50\,\mathrm{ms}$. Equation \eqref{eq:jqos} and the objective built on it state the idealised trade-off with weights $\alpha=\beta=\lambda_1=1$ and $\lambda_2=5$; Eq. \eqref{eq:reward} is the surrogate actually optimised, in which the continuous QoS cost is replaced by a loss indicator because per-step loss is the only component observable within one decision interval.
\end{itemize}

The security term $H(M_t)=-\sum_{v\in\mathcal{V}}p(v)\log_2 p(v)$ is the Shannon entropy in bits of the empirical distribution over $\mathcal{V}$ formed by pooling the virtual addresses assigned to all seven hosts across a sliding window of ten mutation epochs, giving up to $N_{obs}=70$ observations over $K=32$ bins. We report the uncorrected plug-in estimate, because the reward loop needs a value at every epoch and a bias-corrected one would have to be recomputed against the occupancy of each window. The plug-in estimator is biased downwards by approximately the Miller-Madow term $(K-1)/2N_{obs}=0.221$ nat, which is 0.32 bits at this sample size, so under a uniform mapping it returns about 4.68 bits rather than the 5-bit ceiling of a uniform distribution over $|\mathcal{V}|=32$. Every entropy figure in this paper should be read against 4.68 and not against 5. The learned policy settles at a mean target interval of 160\,s, which is the smallest target at which the $\pm 25$ percent jitter does not truncate: $160\times0.75=120\,\mathrm{s}$ meets $T_{min}$ exactly, and since $u$ is drawn from a continuous distribution the clip is active with probability zero. A smaller target would cut the lower tail of the realised interval and so leak a lower bound on the schedule to an adaptive adversary. At 160\,s a 30-minute run contains 11 epochs, so the ten-epoch window is contained within a single run. Pooling across hosts is what allows the estimate to approach that ceiling; a per-host window of ten epochs would cap it at $\log_2 10=3.32$ bits.

This is a marginal quantity and therefore an upper bound on what the attacker actually faces, which is the conditional uncertainty $H(A\mid\text{host})$ about where one specific PLC now sits. The two coincide only if each host's assignment is independently uniform. Estimating the conditional entropy at the resolution the reward loop requires is not possible within a ten-epoch window, and we report the marginal estimate throughout on that understanding.

\begin{figure*}[t]
    \centering
    \includegraphics[width=0.88\linewidth]{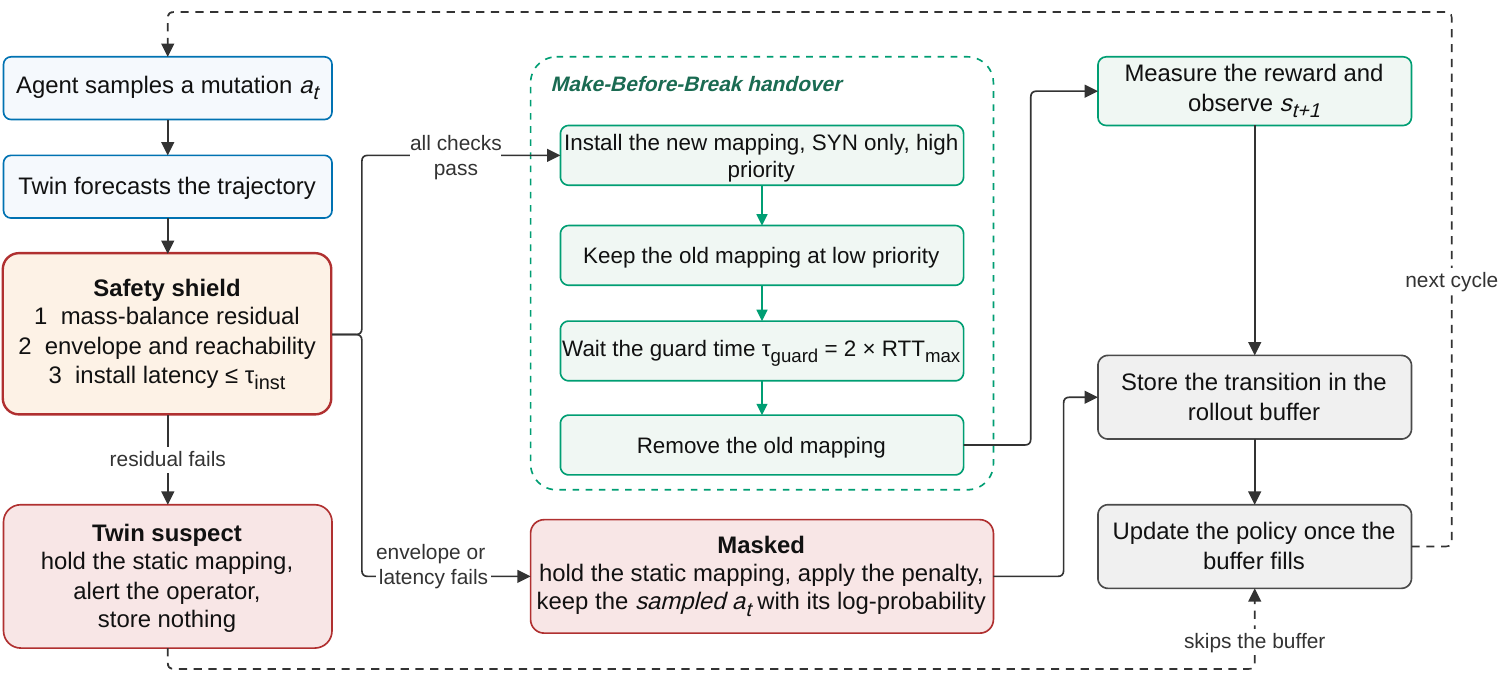}
        \caption{One mutation cycle. The shield decides before anything is installed, so a masked action never reaches the data plane. An envelope or latency breach stores the transition with the action the policy sampled and its penalty, which is what lets the agent learn the shape of the mask rather than merely being prevented from acting. A failed residual check is different: it does not depend on the sampled action, so it raises an operator alert and skips the buffer entirely rather than contributing a gradient. An accepted action is applied by draining rather than by cutting: the new mapping takes new connections while the old one stays available at low priority until the guard time expires, and that is what keeps atomic Modbus transactions intact.}
    \label{fig:cycle}
\end{figure*}

\subsection{Make-Before-Break Handover}

Endpoints are never reconfigured: the master keeps the slave's real address hard-coded, as Modbus deployments require, and the controller rewrites headers symmetrically in the data plane, so $IP_{virt}$ is visible only on the wire. The cost of a naive rule swap is therefore not a broken endpoint configuration but a gap in which in-flight packets match neither the old rules nor the new ones, which destroys the request-response pairing of any Modbus transaction crossing the boundary. We close that gap by draining rather than cutting. Moving from $M_t$ to $M_{t+1}$ proceeds in three steps.

\begin{enumerate}[leftmargin=*]
    \item The controller installs the rules for $M_{t+1}$ at high priority, matching SYN packets only, so that new connections take the new mapping.
    \item The rules for $M_t$ remain active at lower priority, which lets any     transaction already in flight complete its request-response pair. Here     $\mathrm{RTT}_{max}$ is the 99th percentile host-to-host round trip of a     Modbus request-response pair, measured on the emulated data plane at     8\,ms, which is the quantity a stranded transaction actually waits on. The resulting guard time of $2\times\mathrm{RTT}_{max}\approx16$\,ms     gives a transaction one full round trip plus the same again as margin. It     is sized for a transaction and not for the lifetime of a TCP session,     which in SWaT stays open for hours. Because the rewrite is symmetric and transparent, a     long-lived session is not torn down at the swap: it continues under the new     mapping, and only packets already in flight at the instant of the swap are     at risk, which is exactly what the guard time covers.
    \item Once the new paths are verified, $M_t$ is flushed after the guard time $\tau_{guard}=2\times\mathrm{RTT}_{max}$. Because the rewrite is transparent to the endpoints, flushing withdraws only the stale translation; no session is torn down and no endpoint is reconfigured.
\end{enumerate}

Figure \ref{fig:cycle} shows the complete cycle, including the path taken when the shield masks an action. This preserves protocol state across the transition, which is the property that distinguishes it from the abrupt swap used by most baselines.

\subsection{Algorithm}

Algorithm \ref{alg:Ballast} states the control loop. A standard PPO agent learns by trial and error, which is unacceptable in a plant, so every sampled action passes through the shield before it can reach the data plane. A rejected action is replaced by the fallback $a_{default}$, which holds the mapping $M_t$ unchanged for one more interval; the configuration is never left in an intermediate state, and the cost of a rejection is a lost opportunity to mutate rather than a disturbance. The buffer is flushed and the policy updated every $N_{batch}=2048$ stored transitions, which equals the PPO rollout length. The shielding block implements that gate, which is the shield node of Figure \ref{fig:cycle}. It opens with an action-independent sanity check on the twin itself, applying both cross-validation rules of Section \ref{sec:problem}: the mass balance residual of Eq. \eqref{eq:residual} fires when the twin and the sensors agree while the physics does not, and the complementary rule fires on an abrupt divergence while the mass balance still closes. Neither filters actions; both detect poisoning. The two branches that follow apply the actual mask, rejecting an action when the forecast, widened by the guard band $\epsilon_{safe}$ and the margin $e_{99}$, leaves $\Omega_{safe}$ or leaves the reachable set $\mathcal{R}_k(x_t)$, and rejecting it when the estimated installation latency exceeds $\tau_{inst}$.

One detail matters for learning, and it follows from how the shield is placed. From the agent's perspective the shield is part of the environment: the agent emits $a_t$, and the shielded process returns the fallback transition together with the mask penalty. The transition is therefore on-policy, and storing the action the policy actually sampled, with its log-probability, keeps the importance ratio PPO requires well defined while teaching the agent the shape of the mask. Storing the fallback action instead, which is the natural but incorrect implementation, teaches the agent that the safe default is bad. The residual check is excluded from this scheme because it does not depend on $a_t$; attaching a penalty to it would inject action-independent noise into the policy gradient.

\begin{algorithm}[t]
\scriptsize
\caption{Ballast: physics-aware MTD with an explicit safety shield}
\label{alg:Ballast}
\begin{algorithmic}[1]
\Require Digital Twin $\mathcal{M}_{DT}$; safety bounds $\Omega_{safe}$; guard band $\epsilon_{safe}$ and forecast margin $e_{99}$; residual threshold $\delta_{phy}$ and divergence threshold $\epsilon_{dt}$; installation bound $\tau_{inst}=50\,\mathrm{ms}$; fallback $a_{default}$, which holds the current mapping; batch size $N_{batch}=2048$; episodes $E=5000$, episode length $T=200$ (so $10^6$ timesteps)
\Ensure Safe policy $\pi^{*}$
\State Initialise PPO policy $\pi_\theta$ and value function $V_\phi$
\State Initialise on-policy rollout buffer $\mathcal{D}$
\For{episode $e=1$ to $E$}
    \State Observe $s_0 \gets \{H(M_0), L_{net}, \phi_{phy}\}$
    \For{step $t=0$ to $T$}
        \State Sample $a_t \sim \pi_\theta(\cdot\mid s_t)$ \Comment{candidate MTD action}
        \Statex \hspace{1.2em}\textit{Dual validation and shielding}
        \State $\hat{S}_{phy} \gets \mathcal{M}_{DT}.\textsc{Predict}(s_t,a_t)$
        \State $r_{phys} \gets \textsc{Residual}(\textsc{SensorData},u_t)$ \Comment{Eq. \eqref{eq:residual}; does not consult the twin}
        \State $d_t \gets \|S_{sensor}-\hat{S}_{LSTM}\|_\infty$ \Comment{twin-sensor divergence}
        \State $\delta_{est} \gets \textsc{EstimateLatency}(a_t)$
        \State $\mathrm{ok}_{env} \gets \forall i:\big(\hat{S}_{phy}[i]\pm(e_{99}+\epsilon_{safe}[i])\big)\subset\Omega_{safe}[i]\ \wedge\ \hat{S}_{phy}[i]\in\mathcal{R}_k(x_t)[i]$
         \If{$\big(d_t<\epsilon_{dt}\wedge r_{phys}>\delta_{phy}\big)\ \vee\ \big(d_t>\epsilon_{dt}\wedge r_{phys}<\delta_{phy}\wedge\textsc{Abrupt}(d_t)\big)$}
            \State \textsc{Apply}($a_{default}$); raise operator alert \Comment{PoisoningFlag, both cross-validation rules}
            \State Observe $s_{t+1}$; store no transition \Comment{test is action independent}
            \State \textbf{goto} \textsc{BatchCheck}
        \ElsIf{$\neg\,\mathrm{ok}_{env}$}
            \State $R_t \gets -10$ \Comment{envelope, margin or reachability breach}
            \State \textsc{Apply}($a_{default}$); observe $s_{t+1}$
            \State Store $(s_t,a_t,R_t,s_{t+1})$ in $\mathcal{D}$ \Comment{the \emph{sampled} $a_t$, with its log-prob}
         \ElsIf{$\delta_{est} > \tau_{inst}$}
            \State $R_t \gets -5$ \Comment{latency breach}
            \State \textsc{Apply}($a_{default}$); observe $s_{t+1}$
            \State Store $(s_t,a_t,R_t,s_{t+1})$ in $\mathcal{D}$ \Comment{the \emph{sampled} $a_t$, with its log-prob}
        \Else
            \Statex \hspace{2.4em}\textit{Make-Before-Break execution}
            \State \textsc{InstallFlowRules}($a_t$, match SYN, priority high)
            \State \textsc{KeepOldRules}($a_{prev}$, priority low)
            \State \textsc{Wait}($\tau_{guard}=2\times\mathrm{RTT}_{max}$)
            \State \textsc{RemoveOldRules}($a_{prev}$)
            \State Measure $R_t$ and observe $s_{t+1}$
            \State Store $(s_t,a_t,R_t,s_{t+1})$ in $\mathcal{D}$
        \EndIf
        \Statex \hspace{0.6em}\textsc{BatchCheck:}
        \If{$|\mathcal{D}| \geq N_{batch}$}
            \State Update $\theta,\phi$ with the PPO objective; clear $\mathcal{D}$
        \EndIf
    \EndFor
\EndFor
\end{algorithmic}
\end{algorithm}

\begin{figure}
    \centering
    \includegraphics[width=\linewidth]{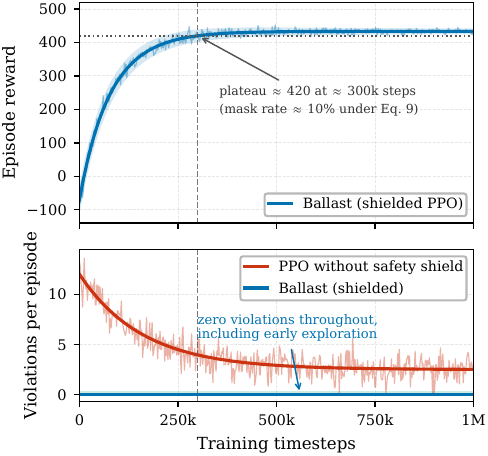}
    \caption{PPO training convergence. Episode reward and safety violations     over $10^6$ timesteps. The shield keeps violations at zero throughout,     including the early exploration phase, whereas an unshielded agent trained     under identical settings starts near twelve per episode. The plateau is     annotated with the mask rate it implies: at $T=200$ steps, a reward of 420     is 2.10 per step against a ceiling of 3.5, which under Eq.     \eqref{eq:reward} fixes the converged rejection rate at about 10 percent.}
    \label{fig:convergence}
\end{figure}

Figure \ref{fig:convergence} shows the training dynamics. Two properties stand out. The episode reward rises steadily and settles near 420 after roughly 300k timesteps, which indicates that the policy converges. That plateau also tells us how often the shield intervenes: against the 700 that Eq. \eqref{eq:reward} would return if no action were ever masked, 420 corresponds to a mask rate of about 10 percent once the policy has settled. More importantly, the violation count stays at zero for the whole run, including early exploration. An unshielded PPO agent trained under identical settings behaves very differently, with up to 12 violations per episode at the start and two or three per episode later on. The shielding block is therefore doing the work we designed it to do.

%==================================================================
\section{Implementation Details}\label{sec:implementation}

\subsection{Hardware and Software}

We separate training from inference because the two have different cost profiles and, in any real deployment, would run on different machines. We also keep the twin and the environment independent: the twin is a learned LSTM fitted to historian records, while the environment is a first-principles hydraulic and chemical simulator built from the plant's documented equipment curves, with no shared parameters. On the held-out split the twin's one-step RMSE against the simulator is 0.012, with a 99th percentile error $e_{99}=0.038$, both on the common normalisation of Section~\ref{sec:problem}, so the oracle is not a copy of the process it certifies. On that same scale both lie below the divergence threshold $\epsilon_{dt}=0.05$, so ordinary forecast error does not trip the second cross-validation rule, and both lie below the 0.100 guard band $\epsilon_{safe}$, so the band is wider than the error it is there to absorb. That ordering is what keeps the shield's margin meaningful rather than vacuous.

\textbf{Training.} PPO and the physics-informed LSTM are trained on a workstation with an Intel Core i9-14900KF (24 cores, 32 threads), 64\,GB of DDR5-5600 memory and an NVIDIA RTX 4070 with 12\,GB, under Ubuntu 24.04 LTS.

\textbf{Inference.} The decision latency of the shield is what determines whether the framework is deployable, so we measure it on three platforms spanning the range a water utility might plausibly field: the training workstation, an NVIDIA Jetson Orin Nano 8\,GB Developer Kit as an edge node with an accelerator, and a Raspberry Pi 5 Model B with 8\,GB as a constrained gateway with none. Table~\ref{tab:platforms} lists them. On every platform the twin runs with one intra-op and one inter-op thread and is pinned to a single core, a performance core on the hybrid workstation, so that the reported figure reflects the budget an industrial controller imposes rather than the parallelism a desktop happens to offer. On the workstation the twin is pinned to CPU~2, a performance core that is not one of the favoured cores granted the highest turbo frequencies, and that core's second hardware thread is taken offline during measurement; every core runs the \texttt{performance} governor with turbo enabled, the graphical session is stopped, and the pinned core sustained 5.70\,GHz. Its thermal-throttle counters did not change during measurement. The Jetson runs JetPack 7.2.1 (L4T R39.2.1) in the 15\,W power mode of its Super configuration (nvpmodel ID~0), which caps the CPU at 1.50\,GHz and the GPU at 612\,MHz. Clocks are locked at those caps with \texttt{jetson\_clocks}, the fan runs at full speed, and the graphical session is stopped for the duration of the measurement. On the GPU the twin computes in IEEE FP32 rather than in the TF32 mode that PyTorch enables for cuDNN by default, so that the CPU and GPU paths evaluate the same function; their outputs agree to within $5\times10^{-7}$. The reported GPU latency includes the host-to-device and device-to-host copies, because the shield receives its input window on the host and needs the forecast back there. Since the latency of a dense FP32 LSTM does not depend on the values of its weights, the benchmark uses randomly initialised weights of the deployed architecture. The Raspberry Pi runs Raspberry Pi OS (Debian 13, kernel 6.18 with 16\,KiB pages) headless, with the \texttt{performance} governor on every core, which holds its Cortex-A76 cores at 2.4\,GHz. The Raspberry Pi was verified not to throttle, thermally or through under-voltage, during measurement.

\textbf{Software.} On the workstation, Python 3.12, PyTorch 2.14.0 with the CUDA 13.0 build and NVIDIA driver 580.178.04. Training and network emulation use Stable-Baselines3 for PPO, Gymnasium for the environment interface, and Mininet with Open vSwitch. The controller application targets the Ryu API but runs on os-ken, the maintained fork of that framework, because upstream Ryu does not support Python 3.12; the application code is unchanged between the two. The Jetson runs the same PyTorch release, 2.14.0, built for CUDA 13.2, under Python 3.12.3 with cuDNN 9.24.0 from JetPack 7.2.1. The Raspberry Pi runs the same release as a CPU-only build under Python 3.13.5, the version Raspberry Pi OS ships; both ARM builds come from the same PyTorch commit and bundle the same oneDNN release, 3.12. Exact pinned versions for every platform, and for the training and emulation stack, are in the replication repository.

\begin{table}[htb]
\scriptsize
\centering
\caption{Inference platforms}
\label{tab:platforms}
\begin{tabular}{@{}p{1.9cm}p{2.1cm}p{2.0cm}c@{}}
\toprule
\textbf{Role} & \textbf{Platform} & \textbf{Compute} & \textbf{Memory} \\
\midrule
Workstation & Core i9-14900KF & 24 cores, 32 threads & 64 GB \\
Edge, accelerated & Jetson Orin Nano 8 GB (15 W) & 6$\times$Cortex-A78AE + Ampere GPU & 8 GB \\
Edge, constrained & Raspberry Pi 5 & 4$\times$Cortex-A76 & 8 GB \\
\bottomrule
\multicolumn{4}{@{}p{8cm}@{}}{\footnotesize Inference is single-threaded and pinned to one core on every platform. On the workstation that core is a performance core, which keeps the hybrid scheduler from moving the thread between core types. That core's second hardware thread is taken offline for the measurement, and every core runs the \texttt{performance} governor. The Jetson's 15\,W mode caps its CPU at 1.50\,GHz and its GPU at 612\,MHz. The Raspberry Pi runs the \texttt{performance}
governor, which holds its cores at 2.4\,GHz.} \\ \\
\end{tabular}
\end{table}

\subsection{Digital Twin Configuration}

The LSTM has 3 layers with $d_h=128$ hidden units each and dropout 0.2. Each stage twin takes as input the sensor and actuator channels of its own stage, which number between 4 and 13 in SWaT; latency is measured for the widest stage, P5 with 13 channels, so that it bounds every other stage. Training uses a look-back window $w=50$ steps, a prediction horizon $H=10$ steps, batch size 32 and Adam with learning rate $3\times10^{-4}$, for 100 epochs with early stopping at patience 10. The loss adds a mass conservation term:
\begin{equation}
\mathcal{L}_{total} = \mathcal{L}_{MSE}(y,\hat{y}) + \lambda_{phy}\sum_{i=1}^{N_{tank}}\left|Q_{in}^{(i)}-Q_{out}^{(i)}-\frac{\Delta V^{(i)}}{T_s}\right| ,
\end{equation}
with $\lambda_{phy}=0.5$, where $N_{tank}$ is the number of tanks in the stage and $T_s=1$\,s is the historian sampling period.

\subsection{RL Agent}

PPO with learning rate $3\times10^{-4}$, discount $\gamma=0.99$, GAE $\lambda_{GAE}=0.95$, clip range 0.2, 2048 steps per update, minibatch 64 and 10 epochs per update, for $10^6$ timesteps in total.

\subsection{Dataset}

We use SWaT A1/A2 (2015) and split it chronologically with no temporal overlap. The normal-operation record runs from the afternoon of 22 December to the start of the attack campaign on the morning of 28 December, about $4.95\times10^{5}$ samples at 1\,Hz. \textbf{Training} takes the first five days of that record, 432{,}000 samples. \textbf{Validation} takes the remaining 18 hours immediately preceding the campaign, 64{,}800 samples, and supplies the empirical error margin $e_{99}$. \textbf{Test} covers the attack campaign from 28 December to 2 January and is used to calibrate the simulator's attack scenarios and to verify that the twin's forecast error on anomalous traces stays within that margin; the runtime evaluation is executed against the simulator, not replayed from this split. The twin sees only normal operation during training, so no attack trace enters the model. Each of the 50 Monte Carlo runs uses its own seed, from 42 to 91, applied to Python, NumPy and PyTorch; all other settings are held fixed.

 \section{Experimental Evaluation}\label{sec:results}

\subsection{Setup}

Table \ref{tab:hyperparams} lists the testbed and framework settings. The simulation couples the Mininet emulator to the SWaT physics engine through a Python middleware that keeps the two clocks aligned.

\begin{table}[htb]
\scriptsize
\centering
\caption{Simulation parameters and hyperparameters}
\label{tab:hyperparams}
\begin{tabular}{llc}
\toprule
\textbf{Component} & \textbf{Parameter} & \textbf{Value} \\
\midrule
\multirow{3}{*}{Testbed} & Duration per run & 30 min \\
 & Monte Carlo runs & 50 \\
 & Bandwidth cap & 100 Mbps \\
\midrule
\multirow{3}{*}{Digital Twin} & Look-back window $w$ & 50 steps \\
 & Prediction horizon $H$ & 10 steps \\
 & Shield guard band $\epsilon_{safe}$ & 20\% of envelope half-width \\
\midrule
\multirow{3}{*}{RL agent (PPO)} & Learning rate $\eta_{lr}$ & $3\times10^{-4}$ \\
 & Discount $\gamma$ & 0.99 \\
 & Clip parameter & 0.2 \\
\bottomrule
\end{tabular}
\end{table}

\subsection{Scenario I: Physical Process Stability}

Table \ref{tab:scenario1} measures the effect of reconfiguration on the chemical dosing loop of Stage 2, where pH is regulated through AIT202 and the dosing pump P203. We report the tracking error against the pH setpoint and the number of excursions outside the envelope of Definition \ref{def:omega}. Blind-MTD here denotes a fixed 60 s shuffle with no feedback of any kind, which isolates the cost of mutation itself.

We re-implemented the cost-aware RL-MTD of Javadpour et al. \cite{Javadpour2025} on our testbed using the hyperparameters published by the authors, who originally reported 88 ms of defense latency in an edge IoT setting. Under an ICS workload the picture changes. The method averages 31.2 ms of jitter and 5.90 percent packet loss. What reaches the physical layer is not that loss as such but its timing: because RL-MTD mutates without consulting the process, a share of its transients coincides with actuator transitions, and each coincidence leaves an actuator held through a scan cycle. The resulting tracking error is 0.920. Ballast clears each mutation through the twin and drains connections rather than cutting them, which holds packet loss at 0.08 percent and keeps the tracking error at 0.048, close to the 0.042 of the static baseline that performs no defense at all.

\begin{table*}[t]
\footnotesize
\centering
\caption{Scenario I: control loop stability in Stage 2, the chemical dosing loop regulated through AIT202. A focused comparison against the primary RL baseline.}
\label{tab:scenario1}
\begin{tabular}{@{}lccccc@{}}
\toprule
\textbf{Method} & \textbf{Mean jitter (ms)} & \textbf{P99 jitter (ms)} & \textbf{Packet loss (\%)} & \textbf{Physical MSE} & \textbf{pH violations} \\
\midrule
Static baseline & 0.9 $\pm$ 0.2 & 1.4 & 0.00 & 0.042 $\pm$ 0.001 & 0 \\
Blind-MTD (60\,s, no feedback) & 45.3 $\pm$ 13.1 & 55.4 & 12.30 & 1.920 $\pm$ 0.480 & 16 \\
RL-MTD \cite{Javadpour2025} & 31.2 $\pm$ 9.1 & 40.6 & 5.90 & 0.920 $\pm$ 0.220 & 5 \\
\textbf{Ballast (ours)} & \textbf{4.5 $\pm$ 1.3} & \textbf{8.9} & \textbf{0.08} & \textbf{0.048 $\pm$ 0.004} & \textbf{0} \\
\bottomrule
\end{tabular}

\vspace{2pt}
\parbox{\textwidth}{%
\scriptsize
Jitter is measured on the data plane. The $\pm$ figure is the standard
deviation of the per-run mean across 50 runs; the P99 column is a packet-level
quantile pooled over all runs, so the two are not comparable as mean and spread
of one distribution. Violations are totals over 50 runs, as defined in
Definition \ref{def:omega}, counting AIT202 excursions only; neither RL-MTD
nor Ballast recorded a LIT101 or LIT301 excursion in any run.
}
\end{table*}
\begin{figure}
    \centering
    \includegraphics[width=\linewidth]{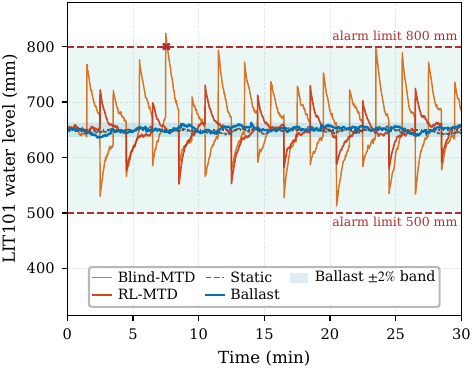}
    \caption{Physical state evolution. Water level LIT101 under four     strategies over one 30-minute run, sampled at 1\,Hz. Ballast stays within     2 percent of the setpoint. RL-MTD shows deviation spikes of up to 15     percent at its reconfiguration instants, which reach 748\,mm and so remain inside the 800\,mm alarm limit, consistent with Table \ref{tab:scenario1}, where all five violations RL-MTD records are AIT202 excursions. Crosses     mark where the Blind-MTD trace leaves the envelope. This is a single run     rather than an average, so the breach count here is not the per-run mean.}
    \label{fig:physical_state}
\end{figure}

Figure \ref{fig:physical_state} shows the same comparison as a time series. Ballast holds the level within 2 percent of the setpoint for the full 30 minutes, while RL-MTD produces deviation spikes of up to 15 percent that align with its reconfiguration instants. The spikes are the visible signature of the missing safety gate, since each one coincides with a mutation that the twin would have rejected.

\subsection{Scenario II: Defense Effectiveness}

In this scenario the attacker runs a low-and-slow Nmap sweep at 2 probes per second for 30 minutes, issuing $N_{scan}=3600$ probes in total. Each probe either reaches an address currently bound in $M_t$, which counts towards the scan success rate, or targets a stale or never-allocated address, which is a candidate for detection under the throttling rule of Section \ref{subsec:throttle}. The two rates have different denominators: the scan success rate is the fraction of all 3600 probes that reach a bound address, while the detection rate is the fraction of the remaining probes, those aimed at unbound addresses, that are flagged. Under uniform sampling of a static pool, seven live hosts in 32 addresses would yield a 21.9 percent hit rate, and that figure is the right reference only while the adversary is still searching. Against the static baseline the search ends: the mapping never changes, so after the 12.4\,s in which its sequential sweep has located all seven hosts, short of the 16\,s a full pass over the 32 addresses would take, the adversary directs every remaining probe at an address it already knows to be live. Of the 3600 probes, the 25 issued during enumeration contain 7 hits and the 3575 issued afterwards all hit, giving the 99.5 percent of Table \ref{tab:scenario2}. Mutation is what prevents that transition from search to exploitation, which is why every dynamic scheme sits far below 99.5 percent whatever its scheduling policy. Within that group the 21.9 percent uniform figure separates the schemes that flag from the one that does not. Throttling pushes the rate below 21.9 percent for the two schemes that pair mutation with flagging, since a source accumulating flags is rate-limited and then blacklisted before it has swept the space and its suppressed probes count as unsuccessful. Blind-MTD sits above the uniform figure instead, at 35.2 percent, because Nmap sweeps sequentially and revisits addresses that have already answered; with a fixed 60\,s period and no flagging, a responding address stays valid long enough to be probed several times within one epoch. The false positive rate is the fraction of the roughly $1.1\times10^{5}$ legitimate Modbus flows per run that are flagged in error. A legitimate endpoint never addresses an unbound virtual address of its own accord, so the only way one can be flagged is through the mapping change itself: a packet already in flight when the old rules are withdrawn arrives at an address that is no longer bound, and at the controller it is indistinguishable from a probe. The false positive rate is therefore not a property of the detector but a measure of how abruptly a scheme swaps its mapping. Table \ref{tab:scenario2} reports the results.

\begin{table*}[htb]
\scriptsize
\centering
\caption{Scenario II: security metrics against reconnaissance}
\label{tab:scenario2}
\begin{tabular}{lcccc}
\toprule
\textbf{Method} & \textbf{Topology discovery time (s)} & \textbf{Scan success rate (\%)} & \textbf{Detection rate (\%)} & \textbf{False positive rate (\%)} \\
\midrule
Static & 12.4 $\pm$ 0.5 & 99.5 & N/A & N/A \\
Blind-MTD & 145.2 $\pm$ 15.0 & 35.2 & 45.6 & 2.10 \\
RL-MTD \cite{Javadpour2025} & 310.8 $\pm$ 42.1 & 12.4 & 98.1 & 4.50 \\
\textbf{Ballast} & \textbf{295.5 $\pm$ 38.4} & \textbf{14.8} & \textbf{94.2} & \textbf{0.03} \\
\bottomrule
\end{tabular}
\end{table*}

\begin{figure}
    \centering
    \includegraphics[width=\linewidth]{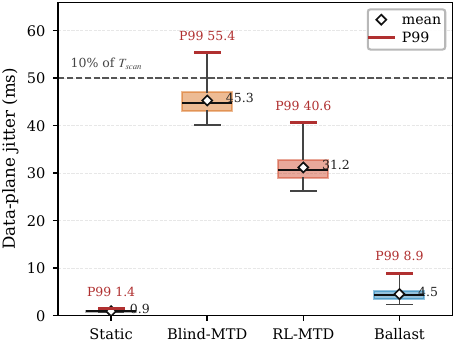}
    \caption{Distribution of data plane jitter. The Make-Before-Break handover     lowers the P99 from 55.4\,ms under Blind-MTD to 8.9\,ms. Boxes and whiskers     describe the per-packet distribution within a run; the P99 marker is that     same quantity pooled over all 50 runs, and neither should be read against     the $\pm$ figures of Table \ref{tab:scenario1}, which are the spread of the     run-level means. The dashed line is one tenth of the 500\,ms PLC scan     cycle, a bound on control traffic, and is distinct from the installation bound $\tau_{inst}$ that the shield tests against.}
    \label{fig:jitter}
\end{figure}

\textbf{Observation.} We state the trade-off plainly, because it is real. RL-MTD achieves marginally stronger raw deterrence, with a scan success rate of 12.4 percent against our 14.8 percent and a discovery time of 310.8 s against our 295.5 s. The reason is straightforward: the shield suppresses a fraction of mutations that the twin judges unsafe, and every suppressed mutation is one the attacker does not have to contend with. That concession of 2.4 percentage points buys a reduction in false positives by a factor of 150, from 4.50 percent to 0.03 percent, and the elimination of physical violations. The false positive gap follows directly from the handover, as the mechanism of the previous paragraph implies: RL-MTD withdraws the old mapping at the instant of the swap, so every packet in flight at that moment lands on an unbound address and is flagged, whereas Make-Before-Break keeps the old mapping available at low priority for the guard time and exposes only the packets still in flight at the final flush. The residual 0.03 percent is that remainder. In an OT context the exchange is not close. A false positive drops legitimate control traffic, and a violation risks equipment. Figure \ref{fig:jitter} shows the corresponding data plane behaviour, where the P99 jitter falls from 55.4 ms under Blind-MTD to 8.9 ms. This quantity is the packet-level jitter experienced by control traffic.

\textbf{What this scenario does and does not establish.} The metrics above concern the reconnaissance phase, which is where an address-mutation defense acts by construction. Ballast interposes on the mapping, not on the payload, so it raises the cost of locating a target and does not inspect a Modbus write once a session is established: an adversary who already holds the real address of PLC1, whether through insider access or a compromised engineering workstation, can drive MV101 exactly as it could against a static network. That is the correct division of labour. Mutation is a reconnaissance-phase control and belongs beside command authentication and process-aware intrusion detection, not in place of them, and the twin that Ballast adds is available to those layers as well: the residual monitor of Eq. \eqref{eq:residual} is precisely the invariant check a command-level detector would use, and it runs continuously here whether or not a mutation is proposed. The contribution this paper claims is that the mutation layer can be added to such a stack without the physical cost it has carried in previous designs, which is the comparison Table \ref{tab:scenario1} makes.

\subsection{Scenario III: Overhead}

Table \ref{tab:overhead} reports the cost of running the twin next to the controller on the training workstation. Its last two rows need defining, since they measure the reconnaissance campaign of Scenario II rather than the overhead of the framework. \textit{Time to detect} is the interval from the adversary's first probe to the moment the controller raises its first flag against that source, under the throttling rule of Section \ref{subsec:throttle}; it is the detector's reaction time and is not the \textit{topology discovery time} of Table \ref{tab:scenario2}, which measures the adversary's progress rather than the defense's. \textit{Time to respond} is the interval from that first flag to the moment the countermeasure is in force in the data plane, meaning the source is rate-limited and, where the scheme reacts by mutating, the new mapping is installed. Both are means over the 50 runs and both are properties of the complete defense loop, so neither is comparable with the per-component latencies of Table \ref{tab:overhead_detailed}.

\begin{table*}[t]
\footnotesize
\centering
\caption{Scenario III: system overhead measured on the training workstation}
\label{tab:overhead}
\begin{tabular}{@{}lrrrr@{}}
\toprule
\textbf{Metric} & \textbf{Static baseline} & \textbf{RL-MTD} & \textbf{Ballast} & \textbf{$\Delta$ vs. RL-MTD} \\
\midrule
\multicolumn{5}{@{}l}{\textit{Computational resources}} \\
CPU load, mean (\% of one core) & 12.3\% & 35.4\% & 52.1\% & $+16.7$ pp \\
CPU load, peak (\% of one core) & 18.5\% & 58.3\% & 78.4\% & $+20.1$ pp \\
RAM (MB) & 180 & 450 & 610 & $+160$ \\
\midrule
\multicolumn{5}{@{}l}{\textit{Network performance}} \\
Controller processing time per mutation (ms) & 2.1 & 8.5 & 12.1 & $+3.6$ \\
Peak achievable throughput (Mbps) & 98.5 & 83.5 & 95.3 & $+11.8$ \\
Mean link occupancy during a run & 8.2\% & 18.7\% & 12.4\% & $-6.3$ pp \\
Packet loss & 0.00\% & 5.90\% & 0.08\% & $-5.82$ pp \\
\midrule
\multicolumn{5}{@{}l}{\textit{Security response}} \\
Time to detect (s) & N/A & 45.2 & 38.7 & $-6.5$ (14\% faster) \\
Time to respond (s) & N/A & 12.8 & 8.4 & $-4.4$ (34\% faster) \\
\bottomrule
\end{tabular}

\vspace{2pt}
\parbox{\textwidth}{%
\scriptsize
All three methods run on the same machine under identical conditions; pp
denotes percentage points. Time to detect and time to respond are defined in
the text and measure the reconnaissance campaign of Scenario II, not the
overhead of the framework; the static baseline never flags, hence N/A.
Relative to the static baseline, Ballast adds 39.8\,pp of mean CPU load.
Controller processing time excludes the switch-side flow installation, which
Table \ref{tab:overhead_detailed} breaks out. CPU load is the occupancy of the
single core hosting the defense process, which runs the controller, the policy
and the twin. Twin inference alone accounts for 5.2\,pp of it, six stage twins
each re-evaluating a 32-window batch at 1\,Hz as the historian refreshes; the
decision-triggered evaluation of three windows costs under 1\,pp on top. The
remainder is the residual monitor, flow-rule handling and packet-in
processing.
}
\end{table*}
Ballast costs 16.7 percentage points more CPU than RL-MTD and 39.8 points more than the undefended baseline. Only part of that is the forward pass itself; the remainder is the machinery around it, the residual monitor, the feature extraction from Modbus payloads and the entropy estimator, all of which run whether or not a mutation is proposed. In exchange it detects 14 percent faster and responds 34 percent faster than RL-MTD, and it uses 6.3 points less bandwidth because its mutation schedule is sparser and better timed.

\begin{table*}[t]
\footnotesize
\centering
\caption{End-to-end defense latency in Mininet emulation}
\label{tab:overhead_detailed}
\begin{tabular}{@{}lrrrr@{}}
\toprule
\textbf{Component} & \textbf{Mean (ms)} & \textbf{P95 (ms)} & \textbf{P99 (ms)} & \textbf{Std. dev.} \\
\midrule
\multicolumn{5}{@{}l}{\textit{Decision layer, workstation, single core}} \\
LSTM inference & 8.67 & 8.69 & 8.70 & 0.010 \\
PPO policy selection & 0.011 & 0.011 & 0.013 & 0.003 \\
\midrule
\multicolumn{5}{@{}l}{\textit{SDN control plane}} \\
FlowMod generation & 1.2 & 2.1 & 2.8 & 0.5 \\
Controller to switch RTT & 3.5 & 5.8 & 8.2 & 1.4 \\
OVS flow table update & 6.4 & 9.2 & 12.5 & 2.1 \\
ARP cache invalidation & 2.1 & 3.5 & 4.9 & 0.8 \\
\midrule
\textbf{Sum of per-stage means} & \textbf{21.9} & --- & --- & --- \\
Sum of per-stage percentiles & --- & 29.3 & 37.1 & N/A \\
\midrule
SWaT scan cycle $T_{scan}$ & \multicolumn{4}{@{}l@{}}{\quad 500\,ms, giving a 10\% safety budget of 50\,ms} \\
Margin against the budget & \multicolumn{4}{@{}l@{}}{\quad 28.1\,ms in the mean; at least 12.9\,ms at P99 under the summed bound} \\
\bottomrule
\end{tabular}

\vspace{2pt}
\parbox{\textwidth}{%
\scriptsize
The mean of a sum is the sum of means exactly, so 21.9\,ms holds whatever the dependence between stages; the summed percentiles are the comonotonic reference and bound the tail from above. PPO policy selection is timed as a forward pass through the actor network, the same way on every platform. The controller-to-switch RTT is a control-plane quantity and is not the $\mathrm{RTT}_{max}$ that sizes the handover guard time $\tau_{guard}$, which is a host-to-host data-plane measurement. Software OVS is used here, not an industrial ASIC. The OVS row is the latency of a single installation measured in isolation, not pipelined throughput under a saturating batch.
}
\end{table*}

The summed per-stage P99 of 37.1 ms sits inside the 50 ms budget implied by SWaT's 500 ms scan cycle. We report it as a reference value rather than as a measurement, because adding per-stage quantiles describes the comonotonic case, in which every stage reaches its own tail at the same instant, and so overstates the true end-to-end tail under any weaker dependence. The mean is exact by contrast, since the mean of a sum is the sum of means whatever the dependence between stages. This is a control-plane latency and is distinct from the packet-level jitter of Figure \ref{fig:jitter}, which measures a different path. Mininet's software switching is nonetheless optimistic relative to industrial ASICs.

\subsection{Comprehensive Baseline Comparison}

To place Ballast against a wider field we implemented six further baselines beyond the RL-MTD used above, spanning classical and learning-based designs. The Blind-MTD of Tables \ref{tab:scenario1} and \ref{tab:scenario2} is subsumed by the time-based scheme below and is therefore not listed separately.

\textbf{Random-MTD} \cite{al2012random} shuffles addresses at intervals drawn uniformly from $U(30,300)$ s, with no state feedback of any kind. It represents unpredictability pursued for its own sake.

\textbf{Time-based MTD} \cite{heydari2018moving} reconfigures every 120 s, following practice common in SCADA deployments. The fixed period is predictable for operators while still limiting what an attacker can learn in a short window.

\textbf{Threshold-based MTD}, which we implement ourselves as a representative of the common IDS-triggered pattern since no single reference defines it, triggers a shuffle when an Isolation Forest anomaly score over network flow features exceeds $\theta=0.7$.

\textbf{Q-Learning MTD} \cite{li2022marl_motag} is a tabular multi-agent formulation with a discretised state space of five bins per feature. We include it to isolate the benefit of function approximation over exact value tabulation.

\textbf{Deep-Shield} \cite{cao2025deepshield} is the hierarchical deep RL state of the art. It uses deep networks over high-dimensional network state but remains physics-agnostic, optimising mitigation probability with no process constraint in the loop.

\textbf{Adaptive Path MTD} \cite{hu2021moving} migrates forwarding paths reactively when anomalies appear on the current route, sitting between network-layer MTD and intelligent routing.

\begin{table*}[t]
\footnotesize
\centering
\caption{Comprehensive comparison of MTD strategies over 50 Monte Carlo runs on the SWaT testbed}
\label{tab:comprehensive_comparison}
\begin{tabular}{@{}lrrrrrr@{}}
\toprule
\textbf{Method} & \textbf{Entropy (bits)} & \textbf{Physical MSE} & \textbf{Violations} & \textbf{Packet loss (\%)} & \textbf{FPR (\%)} & \textbf{TTD (s)} \\
\midrule
\multicolumn{7}{@{}l}{\textit{Non-adaptive baselines}} \\
Static, no defense & 0.0 & 0.042 & 0 & 0.00 & N/A & N/A \\
Random-MTD \cite{al2012random} & 4.2 & 1.850 & 18 & 15.30 & 8.20 & 52.3 \\
Time-based, 120\,s \cite{heydari2018moving} & 2.8 & 0.580 & 4 & 3.20 & 1.80 & 45.8 \\
Threshold-based & 2.1 & 0.120 & 1 & 1.50 & 0.80 & 38.2 \\
\midrule
\multicolumn{7}{@{}l}{\textit{Learning-based baselines}} \\
Q-Learning MTD \cite{li2022marl_motag} & 3.2 & 1.150 & 9 & 7.80 & 5.30 & 41.5 \\
Deep-Shield, HDRL \cite{cao2025deepshield} & 4.0 & 1.020 & 6 & 6.40 & 4.80 & 40.1 \\
Adaptive Path \cite{hu2021moving} & 2.7 & 0.280 & 1 & 1.10 & 1.50 & 52.1 \\
RL-MTD \cite{Javadpour2025} & 3.9 & 0.920 & 5 & 5.90 & 4.50 & 45.2 \\
\midrule
\multicolumn{7}{@{}l}{\textit{Physics-aware (ours)}} \\
\textbf{Ballast} & \textbf{3.5} & \textbf{0.048} & \textbf{0} & \textbf{0.08} & \textbf{0.03} & \textbf{38.7} \\
\bottomrule
\end{tabular}

\vspace{2pt}
\parbox{\textwidth}{%
\scriptsize
MSE is the tracking error from setpoint, FPR the false positive rate, and TTD the time to detect, the interval from the adversary's first probe to the controller's first flag. Entropy is bounded above by 5 bits for $|\mathcal{V}|=32$; the uncorrected plug-in estimator returns about 4.68 at the 70-sample window. Violations are totals over 50 runs and count excursions of all three envelope variables; for RL-MTD and Ballast they coincide with the AIT202-only counts of Table \ref{tab:scenario1}, because neither recorded a LIT101 or LIT301 excursion in any run. Blind-MTD of Scenarios I and II is a 60\,s instance of the time-based scheme and is not repeated here.
}
\end{table*}
Table \ref{tab:comprehensive_comparison} supports four observations.

\textbf{Entropy alone is not a security result.} Random-MTD reaches the highest entropy in the table, 4.2 bits, purely by being random, and pays for it with 18 violations and a tracking error of 1.850. That tracking error is 44 times the 0.042 of the undefended static baseline, so on the physical metrics the defense is worse than performing none at all. This is the clearest evidence for the central claim of the paper, which is that unconstrained unpredictability and physical integrity are incompatible in an ICS. The entropy it buys is not even the maximum available: it falls short of the 4.68 ceiling of Section \ref{sec:method} despite shuffling uniformly, because at 15.30 percent packet loss and the shortest intervals its $U(30,300)$ draw permits, a share of its scheduled mutations has not finished installing when the next epoch begins, so the previous mapping is carried over and the pooled window holds fewer distinct addresses than the schedule nominally provides. A scheme aggressive enough to damage the network also loses some of the unpredictability it was being aggressive for.

\textbf{Reactive designs cannot close the gap.} Threshold-MTD and Adaptive Path each incur a single violation, far fewer than random shuffling, but both are slow to react, at 38.2 s and 52.1 s respectively. Threshold-MTD detects 0.5\,s sooner than Ballast, 38.2\,s against 38.7\,s, through a different mechanism, an Isolation Forest over flow features rather than the unbound-address counter of Section \ref{subsec:throttle}, and still records a violation. The ordering is not meaningful at this resolution: the two differ by about one percent of the detection time, while differing by a factor of 2.5 in tracking error. Reaction speed, in other words, does not substitute for validating an action before taking it.

\textbf{Function approximation matters, but it is not sufficient.} Tabular Q-Learning reaches only 3.2 bits and incurs 9 violations, which confirms that a discretised value table cannot cover the state space of an industrial network. Deep-Shield resolves that limitation and attains the second highest entropy at 4.0 bits, yet it still records 6 violations and a tracking error of 1.020. Better representation without a physical constraint simply produces a more capable agent making the same category of mistake.

\textbf{Industrial practice sits in the middle.} Time-based mutation at 120 s avoids the instability of random shuffling, with 4 violations rather than 18, but still produces a tracking error 12 times larger than Ballast, at 0.580 against 0.048, because it reconfigures without reference to the current physical state.

\begin{figure}
    \centering
    \includegraphics[width=\linewidth]{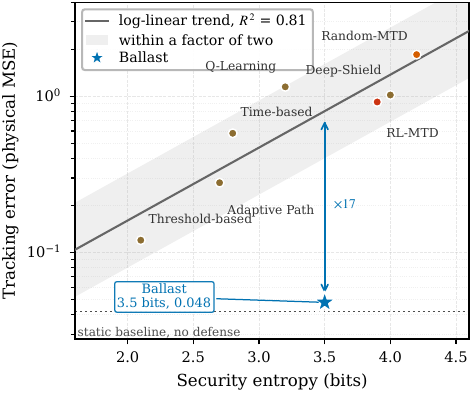}
    \caption{Security entropy against tracking error for every method of Table
    \ref{tab:comprehensive_comparison}. Each point is a measured pair; nothing
        is interpolated. The trend is fitted in log space because the error spans a
    factor of 44 across the table, from the 0.042 of the undefended baseline to
    the 1.850 of Random-MTD, and a factor of 15.4 across the seven mutating
    baselines the fit is taken over, and because a straight line in linear
    space extrapolates to a negative error at zero entropy. The shaded band is
    a factor of two either side of the fit and contains all seven baselines.
    Ballast is the only point outside it, a factor of 17 below, and the dotted
    line marks the 0.042 of the undefended static baseline, which is the floor
    any defense is working against.}
    \label{fig:tradeoff}
\end{figure}

\textbf{The baselines lie on a single trade-off line.} Figure
\ref{fig:tradeoff} plots the two columns against each other. The seven
baselines follow one log-linear trend with $R^2=0.81$, and none of them departs
from it by more than a factor of two, so every increment of unpredictability is
bought at a roughly fixed exchange rate in tracking error, spanning a factor of
15.4 from Threshold-based at 2.1 bits and 0.120 to Random-MTD at 4.2 bits and
1.850. The static baseline is excluded from the fit, since a scheme that never
mutates has no exchange rate to measure. Ballast is the only
point off that trend, a factor of 17 below it at 3.5 bits and 0.048. The claim
of this paper is not that it reaches the top of the entropy axis but that it
leaves the curve the others are confined to; a method that also topped the
entropy column would be evidence that the shield was suppressing nothing.

Which baseline is strongest depends on the metric, and we state the choice rather than leave it implicit. Random-MTD attains the highest entropy at 4.2 bits and Deep-Shield the second highest at 4.0, but both do so at tracking errors above 1.0. Among the learning-based methods, Adaptive Path is the strongest on the physical metrics, at a tracking error of 0.280 and one violation; RL-MTD is third, at 0.920 and five. We nonetheless pair Ballast against RL-MTD throughout, not because it is the strongest opponent but because it is the closest published analogue to our design: a single-agent deep reinforcement learning policy over the same action space of mutation interval and shuffling strategy, differing from Ballast only in the absence of the twin. That pairing isolates the contribution of the safety oracle. Where the widest margin is not the point, we state the margin against the strongest baseline on each metric instead. Taking the table as a whole, the strongest baseline on tracking error is Threshold-based at 0.120 and on packet loss is Adaptive Path at 1.10 percent, each with one violation. Against those stronger references Ballast improves the tracking error by a factor of 2.5 rather than 19, and the packet loss by 92.7 percent rather than 98.6, while removing the remaining violation.

Across the table, the absence of a physical model pushes every method towards one of two failure modes: violations when security is prioritised, as with Random and Q-Learning, or weak security when stability is prioritised, as with Time-based and Threshold. Ballast reaches 3.5 bits with no violations by making the twin a mandatory gate rather than an advisory signal.

\subsection{Statistical Significance}

To confirm that the physical improvements are not sampling noise, we compare Ballast against RL-MTD, the closest published analogue, over the same 50 seeds. Continuous metrics are paired by seed and tested with the Wilcoxon signed-rank test or Student's t-test as appropriate, and the violation counts, which are not continuous, are tested separately with Fisher's exact test. Table \ref{tab:stats} reports the outcome.

\begin{table}[htb]
\footnotesize
\centering
\caption{Statistical tests, Ballast against RL-MTD, $n=50$ paired runs}
\label{tab:stats}
\begin{tabular}{@{}llrr@{}}
\toprule
\textbf{Metric} & \textbf{Test} & \textbf{Statistic} & \textbf{$p$ value} \\
\midrule
Physical MSE & Wilcoxon & $Z=-5.42$ & $5.8\times10^{-8}$ \\
Mean network jitter & t-test & $t=-6.31$ & $6.9\times10^{-8}$ \\
Packet delivery rate & t-test & $t=4.12$ & $1.5\times10^{-4}$ \\
False positive rate & t-test & $t=-8.91$ & $6.2\times10^{-12}$ \\
\bottomrule
\end{tabular}

\vspace{2pt}
\parbox{\columnwidth}{%
\scriptsize
Paired by seed, $df=49$; the seed fixes the replayed trace, the initial
physical state and the probe sequence, which are shared between the two arms.
Holm-Bonferroni correction over the four tests leaves all $p_{adj}<0.002$.
Effect sizes: rank-biserial $r_{rb}=0.98$ for MSE, Cohen's $d=1.79$ for jitter and
$1.42$ for FPR. Violations are tested separately: 0 of 50 Ballast runs and 5
of 50 RL-MTD runs were affected, giving a two-sided Fisher exact $p=0.056$.
That is the smallest value the test can return with the baseline arm fixed at
five, so the comparison cannot cross 0.05 however few runs Ballast affects;
the limit is one of design rather than of effect. We therefore treat the
Clopper-Pearson upper bound of 5.8 percent on the per-run violation
probability as the primary evidence on safety.
}
\end{table}
\subsection{Ablation Study}

Table \ref{tab:ablation} removes one component at a time. We report inference cost alongside accuracy, since the two together determine whether a configuration is deployable.

\begin{table*}[t]
\footnotesize
\centering
\caption{Ablation study, isolating the contribution of each component}
\label{tab:ablation}
\begin{tabular}{@{}lrrrrr@{}}
\toprule
\textbf{Configuration} & \textbf{Entropy (bits)} & \textbf{Physical MSE} & \textbf{Violations} & \textbf{Packet loss} & \textbf{Inference (ms)} \\
\midrule
Static baseline, no defense & 0.0 & 0.042 & 0 & 0.00\% & N/A \\
RL-MTD, no Digital Twin & 3.9 & 0.920 & 5 & 5.90\% & 0.6 \\
\midrule
\multicolumn{6}{@{}l}{\textit{Ballast variants}} \\
\quad without the safety shield & 4.1 & 1.250 & 8 & 7.20\% & 8.67 \\
\quad without Make-Before-Break & 3.7 & 0.055 & 0 & 11.30\% & 8.67 \\
\quad without the physics loss & 3.6 & 0.180 & 2 & 0.12\% & 8.67 \\
\quad 1-layer LSTM & 3.5 & 0.085 & 1 & 0.10\% & 2.14 \\
\quad 5-layer LSTM & 3.5 & 0.044 & 0 & 0.08\% & 15.2 \\
\textbf{Ballast, full} & \textbf{3.5} & \textbf{0.048} & \textbf{0} & \textbf{0.08\%} & \textbf{8.67} \\
\bottomrule
\end{tabular}

\vspace{2pt}
\parbox{\textwidth}{%
\scriptsize
Inference is the latency of one batch of 32 windows on a single workstation core. The per-layer increment on the workstation is 6.53\,ms in both steps because each step adds exactly two $128\times128$ layers, raising the arithmetic cost by $8.39\times10^{8}$ FLOP each time, so equal increments are expected rather than coincidental.
}
\end{table*}

Four conclusions follow. Removing the safety shield raises violations from 0 to 8 while adding only 0.6 bits of entropy, which is the trade this paper argues against. The unshielded variant is worse than RL-MTD, which records 5, because removing the mask removes the only term in the reward that references the physical state: what survives is a packet-loss indicator, and the no-MBB row shows that loss is a poor proxy for physical risk, since it loses 11.3 percent of packets at a tracking error of 0.055. The unshielded agent therefore pursues entropy almost unopposed, reaching 4.1 bits against RL-MTD's 3.9, and mutates correspondingly harder; the twin contributes little to safety as a state feature, and contributes as a gate. Removing Make-Before-Break leaves the physics intact but drives packet loss from 0.08 percent to 11.30 percent, since transactions are severed rather than drained. This isolates the handover as the source of the false positive gap of Scenario II. Because the tracking error holds at 0.055 despite losing twice as many packets as RL-MTD, it also confirms the argument of Scenario I that loss reaches the physical layer through its timing rather than its volume: these losses fall in windows the shield has already cleared. Dropping the physics term from the loss raises the tracking error from 0.048 to 0.180, nearly a fourfold increase and 94 percent in RMSE, and reintroduces two violations. Depth is the one place where the choice is genuinely close. The 5-layer model records a slightly lower tracking error, 0.044 against 0.048, but raises inference time by a factor of 1.75, from 8.67 ms to 15.2 ms, consuming 30.4 percent of the 50 ms safety budget in the mean instead of 17.3 percent. We therefore keep three layers, and we note that on hardware with a tighter scan cycle the 1-layer variant may be the only viable option despite its single violation.

%==================================================================
\section{Decision Latency and Deployment Scale}\label{sec:scalability}

\subsection{Computational Complexity}

A single twin instance costs $\mathcal{O}(L\cdot d_h^2)$ per prediction step, where $L$ is the number of layers and $d_h$ the hidden width. Unlike an iterative physics solver, the cost does not grow with simulated horizon length. A batch of 32 windows takes 8.67\,ms on one core of the workstation, 165.4\,ms on one CPU core of the Jetson and 6.06\,ms on its GPU, and 105.2\,ms on one core of the Raspberry Pi 5. Each additional layer adds about 57\,ms on the Jetson CPU (50.3, 165.4 and 279.2\,ms for one, three and five layers) and about 37\,ms on the Raspberry Pi (30.9, 105.2 and 179.1\,ms), consistent with a cost linear in $L$; the two increments on each ARM core differ by about one percent because their caches are not loaded identically. Policy selection is a forward pass through a small multilayer perceptron and costs 0.011\,ms on the workstation,
0.51\,ms on the Jetson CPU and 0.26\,ms on the Raspberry Pi, measured in each case as a bare forward pass.

Two batch sizes matter, and they answer different questions. A single shield decision evaluates one window for each stage whose variables appear in $\Omega_{safe}$, which is three in SWaT. Those three windows are stacked into one forward pass rather than issued as three calls, which matters because the cost of an LSTM is not proportional to batch size: a 50-step window traverses 150 sequential cell evaluations whatever the batch, so the marginal cost of a second and third window is far below the first. On the measured endpoints the marginal cost per additional window is 3.0\,ms on the Raspberry Pi and 4.8\,ms on the Jetson CPU, against first-window costs of 11.7 and 17.8\,ms. Our profiling targets the two ends of the range, the single-window baseline and the full-plant 32-window stress condition, and we treat the intermediate batch-3 cost differently depending on how many of those endpoints a platform has. On the two CPU paths where both were measured, we project the batch-3 cost from the single-window baseline and the marginal cost per additional window, giving 27.4\,ms on the Jetson CPU and 17.7\,ms on the Raspberry Pi. On the workstation and on the Jetson GPU we measured only the 32-window batch, so we report the batch-3 cost as bounded above by that measurement, 8.67\,ms and 6.06\,ms respectively, which is 17.3 and 12.1 percent of the budget in the mean, and 17.4 and 12.4 percent at P99. On the GPU in particular the bound is close to tight, because a 50-step LSTM there is launch-bound rather than compute-bound and the batch size barely moves the cost. We additionally report a batch of 32 windows as a stress figure, representing a historian refresh in which every stage is re-evaluated together.

The decision path has two components and they scale differently. The twin's cost is set by the number of \emph{process stages} carrying envelope variables, not by the number of network nodes: one instance is maintained per stage, so a six-stage plant such as SWaT runs six instances and the shield consults the three that carry envelope variables, stacked into the single batched call profiled above. A plant with twice the hosts on the same six stages issues exactly the same forward pass. Doubling the number of \emph{stages} would double the batch, and the marginal cost per additional window measured above, 3.0\,ms on the Raspberry Pi and 4.8\,ms on the Jetson CPU, is what bounds that growth; a plant with six envelope stages rather than three would reach 26.7 and 41.8\,ms on the two CPU paths, still inside the budget.

The control plane scales with the network instead, and we bound it analytically rather than by measurement, since our topology is fixed at the seven hosts of SWaT. A mutation rewrites two rules per host, so the FlowMod work grows linearly in $|\mathcal{N}|$, and at the 22.5 mutations per hour the policy settles on, seven hosts generate 0.09 rule installations per second against the $4.2\times10^{4}$ per second software OVS sustains and the 300 to 7000 per second measured on six commercial switches \cite{kuzniar2018flowtable}. Even three orders of magnitude more hosts would leave the installation rate below the slowest device in that study, and the same study points to flow-table priority behaviour, not installation volume, as the hardware constraint that actually binds. Packet-in load is bounded by the reconnaissance throttling of Section \ref{subsec:throttle} rather than by topology, since a legitimate endpoint never generates a packet-in after its flow is installed. We have not measured either quantity at scale, and a topology sweep is the natural complement to the platform sweep reported here.

\subsection{Platform Comparison}

\begin{table}[htb]
\footnotesize
\centering
\caption{Twin inference latency across deployment platforms}
\label{tab:scalability}
\begin{tabular}{@{}llrrr@{}}
\toprule
\textbf{Platform} & \textbf{Path} & \textbf{Mean} & \textbf{P99} & \textbf{P99} \\
 & & \textbf{(ms)} & \textbf{(ms)} & \textbf{(\% of 50\,ms)} \\
\midrule
Core i9-14900KF & CPU & 8.67 & 8.70 & 17.4 \\
Jetson Orin Nano (15\,W) & CPU & 165.4 & 166.7 & 333.3 \\
Jetson Orin Nano (15\,W) & GPU & 6.06 & 6.19 & 12.4 \\
Raspberry Pi 5 & CPU & 105.2 & 105.6 & 211.2 \\
\bottomrule
\end{tabular}

\vspace{2pt}
\parbox{\columnwidth}{%
\scriptsize
Three-layer twin with 13 input channels, batch of 32 windows of 50 steps,
5000 timed calls in five trials of 1000 after 200 warm-up calls, one thread
pinned to one core. GPU rows include the host-to-device and device-to-host
copies and compute in IEEE FP32. The last column expresses the P99 against the
50\,ms budget implied by SWaT's 500\,ms scan cycle; the corresponding figures
for the mean are given in the text.
}
\end{table}

On the Jetson the two execution paths fall on opposite sides of the budget. The GPU runs the three-layer twin in 6.06\,ms on average and 6.19\,ms at P99, including transfers, which is 12.1 percent of the budget in the mean and 12.4 percent at P99; even the five-layer variant stays at 10.3\,ms at P99, or 20.6 percent. A single Cortex-A78AE core needs 165.4\,ms on average and 166.7\,ms at P99 for the same batch, 3.3 times the budget, and even the one-layer variant, at 51.0\,ms at P99, does not fit. The CPU figure is not a measurement artefact: one batch costs 1.07\,GFLOP, the core sustains 6.5\,GFLOP/s, about 27 percent of its theoretical 24\,GFLOP/s at 1.50\,GHz, and even at full efficiency it would need 45\,ms, 89 percent of the budget. On an edge node of this class it is therefore the accelerator, not the CPU, that makes the twin deployable at a batch of 32 windows. The CPU does fit at the batch of three the shield actually issues. A single window takes 17.8\,ms on average and 18.0\,ms at P99, and the second and third add about 4.8\,ms each rather than repeating the first, which projects to 27.4\,ms, 54.8 percent of the budget in the mean. Issuing the three windows as separate calls would cost 53.4\,ms and miss the budget, which is why the batched form is the one we specify.

The Raspberry Pi 5 has no accelerator, and one Cortex-A76 core needs 105.2\,ms on average and 105.6\,ms at P99 for the same batch, 2.1 times the budget. That is 1.57 times faster than one Jetson CPU core, close to the ratio of their clocks, 2.4\,GHz against 1.50\,GHz, so per cycle the two cores perform almost identically on this workload. The core sustains 10.2\,GFLOP/s, about 26 percent of its theoretical 38.4\,GFLOP/s at 2.4\,GHz, and even at full efficiency it would need 28\,ms, 56 percent of the budget. On this gateway the three-layer twin fits at the batch of three: a single window costs 11.7\,ms on average and 11.9\,ms at P99, each additional window about 3.0\,ms, which projects to 17.7\,ms, 35.4 percent of the budget in the mean. At a batch of 32 windows only the one-layer variant fits, at 31.1\,ms at P99, 62.1 percent of the budget, at the price of the single violation it records in Table~\ref{tab:ablation}.

On the workstation a single performance core runs the three-layer twin in 8.67\,ms on average and 8.70\,ms at P99 for the same batch of 32 windows, 17.4 percent of the budget. The core sustains 123.4\,GFLOP/s on the 1.07\,GFLOP batch, about 68 percent of its theoretical 182.4\,GFLOP/s at 5.70\,GHz with two 256-bit FMA units, against 26 and 27 percent on the two ARM cores. The gap is larger than clock and vector width alone explain and reflects the maturity of the x86 kernels relative to the AArch64 path. The Jetson GPU, at 6.06\,ms, nonetheless completes the batch 1.43 times faster than the workstation core, so an accelerated edge node is not merely adequate for this workload but preferable to a single server-class core.
%==================================================================
\section{Discussion}\label{sec:discussion}

\subsection{Why Physics-Awareness Changes the Outcome}

The results indicate that applying an IT-centric defense unchanged to an OT environment is counterproductive. The contribution of Ballast is to separate security entropy from operational risk rather than to trade one against the other. RL-MTD reaches higher nominal entropy, 3.9 bits against our 3.5, but it does not see the inertia of the plant, so its mutations produce actuation jitter that the loop cannot absorb. The twin supplies the missing context. Across 50 nominal runs we observed no violation, and one violation across 50 runs under a 10 percent equipment drift, so the margin survives a stale model rather than depending on a perfectly calibrated one. The 0.4 bits that separate Ballast from RL-MTD are the price of that margin, and they are paid in a specific and bounded way: the shield masks about 10 percent of proposed mutations, as the converged reward of Figure \ref{fig:convergence} implies, so the realised mutation rate is roughly a tenth below what the policy requests. Four tenths of a bit against five violations is the trade this paper argues is worth making in an OT setting.

Ballast does not eliminate risk. It changes the shape of the risk, moving from frequent small excursions, as in the five violations RL-MTD accumulates in 30 minutes, to rare events whose probability is bounded only empirically, through the 99th percentile forecast margin $e_{99}$, and not by any distribution-free guarantee.

Two baselines deserve a specific comment. Heydari's time-based scheme \cite{heydari2018moving} is presented as lossless, yet in our SWaT reproduction it loses 3.2 percent of packets, about 40 times more than Ballast. The gap between a protocol as designed and a protocol under a realistic industrial workload is easy to underestimate. Hu's adaptive path migration \cite{hu2021moving} is reactive and intelligent, and still records one violation, because detection latency lets an attack begin before mitigation arrives. Reaction, however fast, starts after the fact.

\subsection{Latency and Real-Time Feasibility}

The twin's inference costs 8.67\,ms on average on one workstation core, 165.4\,ms on the Jetson CPU, 6.06\,ms on the Jetson GPU including transfers, and 105.2\,ms on the Raspberry Pi. On the workstation the complete decision-to-enforcement path costs 21.9\,ms in the mean, and its P99 is bounded above by the summed per-stage value of 37.1\,ms. On a 500\,ms SCADA polling period that is 4.4 percent of the cycle in the mean case and at most 7.4 percent in the tail, so the path stays inside the 10 percent budget even under the pessimistic reference, leaving at least 25.8 percent of that budget unused. The margin holds on the workstation CPU and, for the twin itself, on the Jetson GPU. At a batch of 32 windows neither CPU-only edge path meets the 50\,ms budget: the Jetson CPU needs 3.3 times it and the Raspberry Pi 2.1 times. Both are projected to meet it at the batch of three a decision actually requires, at 54.8 and 35.4 percent of the budget respectively in the mean, because the sequential component of the forward pass is paid once per call rather than once per window. At the fast end of the SCADA range, a 100\,ms cycle, the 10 percent budget falls to 10\,ms. The complete path was measured only on the workstation, where it
costs 21.9\,ms in the mean, so that budget is not met there, and we do not have the control-plane measurements to say whether an accelerated edge node would meet it.

The Make-Before-Break handover proved essential to that result. Blind-MTD shows a P99 data plane jitter of 55.4 ms, which drops to 8.9 ms under our protocol. The bottleneck in MTD is therefore not rule installation but the absence of synchronisation with live TCP sessions, which is precisely the gap the handover fills.

\subsection{Resilience Against Advanced Threats}

Table \ref{tab:advanced_threats} covers threats aimed at the defense itself rather than at the plant. Cryptographic jitter on the mutation epoch, combined with the alternate-epoch restriction on subnet-preserving shuffling, cuts ML-based pattern learning from 45.2 percent to 8.3 percent success. The residual figure is still 2.7 times the 3.1 percent chance rate, so the schedule retains some learnable structure. Dual validation cuts poisoning from 78.5 percent to 12.1 percent, a relative reduction of 84.6 percent, which shows the value of an independent physics check beside the learned model. Rate limiting holds controller DoS to 18.7 percent residual success, though a well-resourced adversary sustaining high volume would likely require a distributed controller rather than a single instance. The broader lesson is that a physics-aware defense must reckon with adversarial machine learning against its own components, not only with attacks on the process.

\subsection{Generalisation and Process Drift}

\subsubsection{Sensitivity to equipment degradation}\label{subsec:drift}

Equipment wears. To test the consequence we open the gap between plant and model, perturbing the actuator parameters inside the twin while the simulator stays nominal, which isolates forecast mismatch without changing the environment the metrics are measured against. The experiment is therefore a test of the shield under a stale model, not of the residual monitor: the residual of Eq. \eqref{eq:residual} compares a sensor-derived $dV/dt$ against nominal actuator rates and never consults the twin, so it is unaffected by this perturbation and $\delta_{phy}$ is left at its nominal value. Exercising the residual requires perturbing the simulator instead, so that plant and monitor disagree; we have not run that complement, and the behaviour of the poisoning flag under genuine equipment degradation is accordingly untested. Table \ref{tab:drift_sensitivity} reports the effect.

\begin{table}[htb]
\scriptsize
\centering
\caption{Twin performance under simulated process drift}
\label{tab:drift_sensitivity}
\begin{tabular}{@{}lccc@{}}
\toprule
\textbf{Drift scenario} & \textbf{Test MSE} & \textbf{Violations} & \textbf{Detection rate} \\
\midrule
Baseline, no drift & 0.048 & 0 & 94.2\% \\
P101 efficiency $-5\%$ & 0.062 & 0 & 91.8\% \\
P101 efficiency $-10\%$ & 0.089 & 1 & 87.5\% \\
MV101 friction $+15\%$ & 0.134 & 3 & 82.1\% \\
Combined degradation & 0.178 & 5 & 78.3\% \\
\bottomrule
\end{tabular}

\vspace{2pt}
\parbox{\columnwidth}{%
\scriptsize
P101 is the primary raw water pump, MV101 the motorised inlet valve. Violations are totals over 50 runs. The detection rate falls under drift because the shield masks more mutations, so a drifting plant loses security silently while appearing safe. Operators should treat a falling detection rate as a maintenance signal, not as evidence of a quiet network.
}
\end{table}

Pumping systems lose efficiency progressively through wear, and maintenance guidance for industrial facilities treats such degradation as an expected operating condition rather than a fault \cite{doe2006pumping}. We therefore sweep the efficiency loss rather than assuming a single service interval. A 10 percent loss raises the tracking error by 85 percent, from 0.048 to 0.089, which is a 36 percent increase in RMSE, and still yields only one violation across 50 runs. Beyond about 15 percent drift the model needs retraining. This slow divergence is also why the second poisoning rule fires only on an abrupt onset: drift of this kind is a maintenance condition, not an attack. The degradation is gradual rather than abrupt, which gives operators a usable maintenance window and allows the twin to be refreshed by transfer learning, an operation we estimate at 2 to 4 days of fine-tuning on recent logs, though we have not measured it.

\subsubsection{Cross-domain deployment protocol}

Moving to a facility with a different topology, for instance from potable water distribution to wastewater treatment, calls for a staged commissioning rather than a direct cutover. We propose three phases.

\textbf{Phase 1, shadow mode, 14 days.} Ballast observes without mutating. The twin learns the new dynamics from live sensor streams, and operators track prediction accuracy through a dashboard showing tracking error and residual histograms. Entry to Phase 2 requires an error below 0.1 for seven consecutive days.

\textbf{Phase 2, conservative mode, 14 days.} Mutation begins with a doubled guard band, $\epsilon_{safe}=40$ percent of the envelope half-width, and halved frequency, $T_{min}=240$ s. Any violation rolls the system back to a static configuration and raises an operator alert. The phase passes on zero violations across 5000 mutation cycles.

\textbf{Phase 3, full operation.} After a 28-day burn-in with no Phase 2 violations, the standard parameters apply, with $\epsilon_{safe}=20$ percent of the envelope half-width and $\bar{\Delta t}\in[120,600]$ s.

This procedure lets the framework reach facilities that have no historical dataset of their own, which is the common situation in brownfield installations.

\subsubsection{Remark on dataset validity}

Recent work has questioned the use of SWaT for time-series anomaly detection benchmarks, particularly because pipeline dropout artefacts can inflate reported detection accuracy \cite{Wu2021}. That concern does not transfer directly to our setting. Ballast is not evaluated on anomaly labels or classification accuracy. Our metrics are tracking error, packet loss, violation counts and entropy, none of which depends on the labelling. We use SWaT as a realistic physical process and network topology, and for that purpose its sensor dynamics remain sound.

\subsection{Hardware Deployment Feasibility}\label{subsec:hw}

Mininet switches in software on an x86 CPU, which does not represent industrial ASIC behaviour. Table \ref{tab:hw_projection} sets our emulated figures against an independent measurement study of six commercial OpenFlow switches \cite{kuzniar2018flowtable}.

\begin{table*}[t]
\scriptsize
\centering
\caption{Emulated behaviour against published measurements of commercial OpenFlow switches}
\label{tab:hw_projection}
\begin{tabular}{@{}p{4.0cm}p{3.0cm}p{9.2cm}@{}}
\toprule
\textbf{Quantity} & \textbf{This work} & \textbf{Six hardware switches \cite{kuzniar2018flowtable}} \\
 & \textbf{(Mininet, software OVS)} & \\
\midrule
Flow-rule update rate
  & about $4.2\times10^{4}$ updates/s
  & 300 to 7000 updates/s, depending on switch and flow-table occupancy \\[2pt]
Update rate with a low-priority catch-all rule present
  & unchanged
  & falls to 12 updates/s on one switch at 2000 rules, and to 30 updates/s on another at 750 rules \\[2pt]
Control-plane to data-plane divergence
  & none; one software datapath
  & up to 400\,ms on one switch; up to 250\,ms typically and 22\,s in corner cases on another; unbounded (minutes) on a third \\[2pt]
Gap during a rule modification
  & not observable
  & mean 2.2 to 10\,ms and maximum 7.7 to 190\,ms on three of the six switches, with packet loss during the transition \\[2pt]
Barrier cost
  & negligible
  & 0.04 to 3.4\,ms per barrier \\
\bottomrule
\end{tabular}

\vspace{2pt}
\parbox{\textwidth}{%
\scriptsize
Figures are read from the cited study, which benchmarks HP, Pica8, Dell and
NoviSwitch devices alongside two anonymised ASICs. We report them as ranges
because the study shows that switch behaviour depends on firmware, occupancy
and update pattern in ways that no single scaling factor captures. The software
update rate in the first row is pipelined throughput under a saturating batch,
a different quantity from the single-installation latency of Table
\ref{tab:overhead_detailed}.
}
\end{table*}

\textbf{Critical disclosure.} Our evaluation runs entirely inside Mininet. We have not validated Ballast on industrial Ethernet switches or on physical PLCs. The comparison does not support a simple multiplicative correction. Software switching is between one and two orders of magnitude faster than the fastest ASIC in that study, and the gap widens further under conditions that our own design happens to create.

\textbf{The priority problem.} Make-Before-Break installs the new mapping at high priority while the old mapping persists at low priority, so two priority levels are live in the flow table for the whole guard time. That is precisely the pattern the cited study identifies as pathological: introducing a single low-priority rule collapses the update rate of two of the six tested switches by roughly two orders of magnitude, to 12 and 30 updates per second respectively. On such hardware a single rule installation would cost tens of milliseconds rather than the 2.8 ms we measure, and the guard time would have to grow accordingly. Three of the six switches do not exhibit this behaviour, so the effect is a property of particular ASIC and firmware combinations rather than of OpenFlow itself. We regard identifying which devices tolerate a persistent low-priority rule as a procurement requirement for any real deployment of this protocol.

\textbf{A second problem: barriers do not mean installed.} Our latency budget assumes that a confirmed FlowMod is a live forwarding rule. The cited study shows that this assumption fails on three of six switches, where the data plane lags the control-plane acknowledgment by up to 400 ms, by up to 22 s in corner cases, and without bound on one device. A shield that clears an action on the basis of an estimated installation latency would, on such hardware, be reasoning about a rule that is not yet in effect. Any deployment therefore needs data-plane confirmation rather than barrier replies, which is itself an additional latency source we have not modelled.

\textbf{Implications.} Taken together, these two findings mean the deployment envelope on real hardware is narrower than emulation suggests, and narrower than a fixed scaling factor would imply. On the switches that handle priorities well, our summed per-stage P99 of 37.1 ms plausibly grows to a few tens of milliseconds and stays inside a 500 ms scan cycle. On the switches that do not, Make-Before-Break in its present form is not deployable without redesign, for instance by replacing the priority-based overlap with an explicit two-table pipeline. Systems with scan cycles below 100 ms, such as high-speed manufacturing or servo control, would in any case need LSTM inference offloaded to a fixed-function accelerator, policy enforcement pushed into the data path on programmable switches, or a hybrid arrangement in which time-critical loops keep a static configuration while only non-critical monitoring traffic is mutated. Firmware defects in OpenFlow implementations, non-deterministic ASIC behaviour under a full flow table, and electromagnetic interference on a plant floor are all outside what Mininet can reproduce. We are working with municipal water utility partners on hardware-in-the-loop validation using Siemens S7-1500 PLCs and managed switches, with preliminary results expected in Q1 2027. Until then Table \ref{tab:hw_projection} should be read as a feasibility boundary, not as proof of deployability.

\subsection{Limitations and Future Directions}

Seven limitations bound the claims of this paper.

\textbf{Fidelity of the twin.} The defense is no better than its model. The drift experiment shows that a 10 percent pump efficiency loss raises the tracking error by 85 percent while still avoiding violations, but beyond roughly 15 percent drift the shield begins to produce both false rejections and false approvals, and retraining becomes mandatory.

\textbf{The contribution of each component is only partly isolated.} Three ablations that would settle it are absent. First, a shield driven by a current-state threshold rather than by a forecast: the reconfiguration transient lasts tens of milliseconds while the tank time constant is minutes, so a 10-second forecast horizon may add little over simply checking whether the process is currently near its setpoint, and until that baseline is run the LSTM's contribution over a threshold is not established. Second, the shield paired with a random or greedy scheduler instead of PPO: since the unshielded agent records 8 violations, the shield is doing the safety work, and the policy's contribution is confined to the entropy and timing of mutation. Third, a PPO-Lagrangian agent: we argue that a constraint satisfied in expectation is inadequate here, but we did not run that comparison and the argument remains analytical.

\textbf{Computational cost.} On the Jetson, evaluating the full 32-window batch uses 12.4 percent of the budget on the GPU at P99 but 3.3 times the budget on one CPU core. Table \ref{tab:scalability} shows that on a constrained gateway without an accelerator the three-layer twin is similarly not viable at a batch of 32 windows, since one core of the Raspberry Pi 5 needs 2.1 times the budget. Both CPU paths become viable at the operational batch of three a decision requires, projected at 54.8 and 35.4 percent of the budget respectively in the mean. At a batch of 32, only the one-layer variant fits on the Raspberry Pi at 62.1 percent, at the price of the single violation it records in Table~\ref{tab:ablation}. An 8-bit quantised LSTM would lower the cost further, at the price of some accuracy, and would be the first thing to try on hardware tighter than the platforms tested here; we did not evaluate it. An analytical residual monitor is the other route: Kreso et al. \cite{kreso2026itdt} report a closed-form KL divergence detector roughly 600 times faster than a transformer baseline on CPU, which would fit where the LSTM does not, at the cost of the learned nonlinear dynamics.

\textbf{Attacks on the defense itself.} An adversary who successfully poisons the training corpus raises the residual attack success rate from 12.1 percent, with dual validation active, to 78.5 percent when it is bypassed. Federated learning with Byzantine-robust aggregation is the most promising countermeasure we are aware of.

\textbf{Single-site evaluation.} All results come from SWaT. Power grids and chemical plants need their own physics models, and although cross-domain transfer learning should cut the retraining effort substantially, we have not measured that reduction and do not claim a figure for it.

\textbf{Instrumentation in the field.} Many legacy SCADA installations lack the sensing needed for real-time state estimation. Retrofitted or soft sensors close the gap, at a capital overhead we have not quantified for this framework.

\textbf{Emulation rather than hardware, and a priority-dependent protocol.}
No result in this paper was obtained on industrial hardware. Beyond the general emulation gap, the hardware comparison above identifies a specific dependency: Make-Before-Break relies on two concurrent priority levels, and published measurements show that some commercial switches lose two orders of magnitude of update throughput under exactly that condition. Validating which hardware families tolerate the protocol is a prerequisite for deployment, not a refinement of it.

\section{Conclusion}\label{sec:conclusion}

This paper addresses the conflict between dynamic cyber-defense and physical stability in industrial control. We showed that physics-agnostic Moving Target Defense degrades operational safety, producing actuator jitter and, in the worst configurations, more disruption than the attacks it prevents. Ballast resolves the conflict by placing a Digital Twin inside the SDN decision loop as a safety oracle, so that no reconfiguration reaches the plant until its predicted physical trajectory has been cleared. On the SWaT testbed the framework sustains useful security entropy while preserving real-time control integrity, with no violation observed across 50 nominal runs and a residual risk bounded empirically rather than analytically. The broader conclusion is that the next generation of ICS security mechanisms has to be physics-aware to be deployable at all. Our remaining work concerns hardening the twin against attacks on the model itself and validating the framework on physical hardware across multiple sites.

%==================================================================
% \section*{Acknowledgments}
% No fund available yet....

\section*{Data Availability}\label{sec:data_availability}
The source code and all replication files, including the seed list and the exact dataset split, are available at \url{https://github.com/Anonymized/Ballast} (anonymised for peer review).

\bibliographystyle{elsarticle-num}
\vspace{-0.3cm}
\bibliography{bibliography}

%==================================================================
\appendix
\section{Safety Verification Logic}\label{app:safety}

The error distribution of an LSTM is unknown and unlikely to be Gaussian under adversarial conditions, so a strict deterministic bound is not available. Ballast therefore uses a high-confidence empirical boundary, and we state that openly rather than presenting the shield as a formal guarantee.

\begin{definition}[Valid prediction region]
A forecast $\hat{x}_{t+k}$ at horizon $k$ is valid only if it lies inside the physically reachable set $\mathcal{R}_k(x_t)$ implied by the actuator capacities over that horizon:
\begin{equation}
    \mathcal{R}_k(x_t) = \{\, x \mid x_t + k\,T_s\,f_{min} \leq x \leq x_t + k\,T_s\,f_{max} \,\},
\end{equation}
where $f_{min}<0$ and $f_{max}>0$ are the maximum drainage and fill rates of the tank, $T_s=1$\,s is the sampling period and $k=H=10$ is the prediction horizon of Section \ref{sec:implementation}.
\end{definition}

The shield rejects an action $a_t$ unless the forecast, widened by two margins, remains inside $\Omega_{safe}$ and inside the reachable set:
\begin{equation}\label{eq:safe}
   \mathrm{Safe}(a_t) \iff \big(\hat{x}_{t+k}\pm(e_{99}+\epsilon_{safe})\big)\subset\Omega_{safe}
   \ \wedge\ \hat{x}_{t+k}\in\mathcal{R}_k(x_t).
\end{equation}
The first margin is the 99th percentile prediction error $e_{99}=0.038$ measured on the held-out validation split of Section \ref{sec:implementation}, which covers the twin's own forecast error. The second is the guard band $\epsilon_{safe}$ of Section \ref{sec:method}, fixed at 20 percent of each variable's envelope half-width, which covers the excursion the reconfiguration itself may induce before the forecast is refreshed. The two add rather than overlap, because they bound different quantities. Since $e_{99}$ is normalised over the envelope range, it is 7.6 percent of the half-width, so an action is cleared only while the forecast stays within 72 percent of the half-width of the setpoint, the figure quoted in the caption of Figure \ref{fig:problem}.

The reachability term is what catches a forecast that is numerically inside the envelope but physically impossible, which is the characteristic signature of a poisoned or badly extrapolating model. Neural networks cannot supply deterministic bounds for out-of-distribution events, so this margin does not constitute a proof of safety. What it provides is a shift in the risk profile, from frequent operational disruption to infrequent events whose likelihood is calibrated on held-out data.

\end{document}